\documentclass[usenatbib,longnamesfirst]{mn2e}
\RequirePackage{graphicx}
\usepackage{hyperref} 

\begin{document}
\title[Akari mapping of pre-stellar cores]{The initial conditions of isolated star formation -- IX. Akari mapping of an externally heated pre-stellar core}
\author[Nutter, Stamatellos \& Ward-Thompson]{D. Nutter\thanks{E-mail: 
David.Nutter@astro.cf.ac.uk}, D. Stamatellos, D. Ward-Thompson\\
School of Physics and Astronomy, Cardiff University, Queens Buildings, 
Cardiff, CF24 3AA}

\maketitle

\begin{abstract}
We present observations of L1155 and L1148 in the Cepheus molecular cloud, taken using the Far Infrared Surveyor (FIS) instrument on the Akari satellite. We compare these data to submillimetre data taken using the Submillimetre Common-User Bolometer Array (SCUBA) camera on the James Clerk Maxwell Telescope (JCMT), and far-infrared data taken with the imaging photo-polarimeter (ISOPHOT) camera on board the Infrared Space Observatory (ISO) satellite. The Akari data cover a similar spectral window and are consistent with the ISO data. All of the data show a relation between the position of the peak of emission and the wavelength for the core of L1155. We interpret this as a temperature gradient. We fit modified blackbody curves to the spectral energy distributions at two positions in the core and see that the central core in L1155 (L1155C) is approximately 2 degrees warmer at one edge than it is in the centre. We consider a number of possible heating sources and conclude that the A6V star BD+67 1263 is the most likely candidate.  This star is at a distance of 0.7 pc from the front of L1155C in the plane of the sky. We carry out radiative transfer modelling of the L1155C core including the effects from the nearby star. We find that we can generate a good fit to the observed data at all wavelengths, and demonstrate that the different morphologies of the core at different wavelengths can be explained by the observed 2 degree temperature gradient. The L1148 core exhibits a similar morphology to that of L1155C, and the data are also consistent with a temperature gradient across the core. In this case, the most likely heating source is the star BD197053. Our findings illustrate very clearly that the apparent observed morphology of a pre-stellar core can be highly dependent on the wavelength of the observation, and that temperature gradients must be taken into account before converting images into column density distributions. This is important to note when interpreting Akari and Spitzer data and will also be significant for Herschel data.
\end{abstract}


\section{Introduction}\label{intro}

Stars are known to form in the molecular cloud cores that make up the densest parts of the interstellar medium (ISM). The initial conditions for the star-formation process are represented by pre-stellar cores \citep{1994MNRAS.268..276W,2000prpl.conf...59A}. A pre-stellar core is a gravitationally bound fragment of cloud that has not yet formed a hydrostatic object (i.e. a protostar) at the centre \citep{2007prpl.conf...33W}. If we are to understand the processes by which these objects are formed, and the manner in which the go on to form stars, it is important to obtain observational data showing the density structure of these initial conditions.

Due to their location, embedded in the centres of molecular clouds, pre-stellar cores can only be observed in emission at far-infrared (far-IR) and submillimetre (sub-mm) wavelengths \citep{1994MNRAS.268..276W}, and in absorption at mid-IR wavelengths \citep{2000A&A...361..555B}. In the far-IR and sub-mm, we observe the thermal emission of cold dust grains. Pre-stellar cores were first identified observationally by \citet{1994MNRAS.268..276W} in the sub-mm, using the James Clerk Maxwell telescope (JCMT). Sub-mm observations are ideal for observing pre-stellar cores, in that they are sensitive to the coldest, densest parts of molecular clouds \citep[for more details, see][]{2007prpl.conf...33W}. In addition, these observations can be carried out from the ground, which means that we can take advantage of the large collecting area and high angular resolution of large ground-based telescopes such as the JCMT. In this series of papers we are studying the detailed properties of pre-stellar cores \citep{1994MNRAS.268..276W,1996A&A...314..625A,1999MNRAS.305..143W,2001MNRAS.323.1025J,2002MNRAS.329..257W,2005MNRAS.360.1506K,2007MNRAS.375..843K,2008MNRAS.391..205S}.

However, we cannot accurately determine the temperature of the emitting grains from sub-mm observations alone, as these observations all lie on the Rayleigh-Jeans slope of the spectral energy distribution (SED). This temperature information is crucial as it is necessary for converting the observed flux density into a mass distribution. In order to obtain this temperature information, we need observations shortward of the peak of the SED, i.e. in the far-infrared ($\lambda < 200 \mu$m). These wavelengths are only obtainable from space-based telescopes. In this study we compare observations of a pre-stellar core in the Cepheus Flare molecular cloud, using both ground-based sub-mm observations, and space-based far-IR observations from Akari.

The Cepheus Flare covers a large region of the sky, and contains a number of local ($< 500$~pc) star-forming clouds, as well as more distant OB associations. Many of these clouds were originally classified by \citet[][]{1962ApJS....7....1L} based on the obscuration of background starlight. The Cepheus Flare was studied in more detail by \citet[][]{1997ApJS..110...21Y} using the molecular tracer $\rm ^{13}CO$ to derive column densities, sizes and masses for the component clouds. Distance estimates for the nearby Cepheus Flare clouds range from $\sim200$ to $\sim400$ pc \citep[][]{1968AJ.....73..233R,1969MmSAI..40...75V,1981ApJS...45..121S,1992BaltA...1..149S,1993A&A...272..235K,1997A&A...324L..33V,1998ApJS..115...59K,2000MNRAS.319..777K,2004A&A...425..133B}, a spread which is comparable to the linear separation of the clouds on the plane of the sky. The Cepheus Flare region has been extensively reviewed by \citet{2008arXiv0809.4761K}.

Together with the majority of star-forming clouds within 500~pc of the sun, the Cepheus Flare forms part of a larger structure called the Gould Belt, which is a ring of molecular clouds and OB associations, approximately 700~pc across. The Cepheus Flare clouds are located at Galactic longitudes $110-115^\circ$, and latitudes $10-20^\circ$.

In this work, we study the L1155 and L1148 clouds in Cepheus using the Far Infrared Surveyor instrument on the Akari telescope. L1155 and L1148 are part of a sub-group called the L1147/L1158 complex, which also contains L1147, L1152, L1158, and the well studied outflow source, L1157. The sub-group was determined to be at a distance of ${\rm 325 \pm 13 pc}$ \citep[][]{1992BaltA...1..149S}, based on a plot of extinction vs. distance for a number of stars along the line of sight to the sub-group.

\section{Observations}\label{observations}
\subsection{Akari Data}\label{obs_akari}
Akari is a space-based telescope operating across the entire infrared band, from 2.4~$\mu$m to 160~$\mu$m \citep[][]{2007PASJ...59S.369M}. The satellite was launched and is operated by the Japan Aerospace Exploration Agency (JAXA). Its primary mission is to make an all-sky survey with greater spatial resolution and spectral coverage than the data produced by the IRAS telescope \citep[][]{1984ApJ...278L...1N}. The satellite is in a circular sub-synchronous polar orbit, allowing continuous scanning of the sky whilst avoiding pointing at the sun or the earth. In addition to the all-sky survey, a limited number of 'observatory-type' pointed observations are possible, within these pointing constraints.

The data described here were all taken using the Far Infrared Surveyor (FIS) instrument \citep[][]{2007PASJ...59S.389K}. This camera has four photometric bands: the N60 band is centred at 65~$\mu$m and has a bandwidth of 22~$\mu$m; the WIDE-S band is centred at 90~$\mu$m and has a bandwidth of 38~$\mu$m; the WIDE-L band is centred at 140~$\mu$m and has a bandwidth of 52~$\mu$m; and the N160 band is centred at 160~$\mu$m and has a bandwidth of 34~$\mu$m. 

The N60, WIDE-S, WIDE-L and N160 detector arrays have 2$\times$20, 3$\times$20, 3$\times$15 and 2$\times$15 pixels respectively. The pixel scales are 26.8 arcsec for the two short-wavelength arrays, and 44.2 arcsec for the two long-wavelength arrays. The point spread function (PSF) at the  four wavelengths are 37, 39, 58 and 61 arcsec respectively. The calibration accuracy is 20\% for the N60 and WIDE-S bands, 30\% for the WIDE-L band, and 40\% for the N160 band.

The FIS01 observing mode was used, which scans the arrays across the sky at an angle of 26.5$^\circ$ to the short axes of the arrays. The arrays are scanned at a relatively slow scan-speed of 15 arcsec s$^{-1}$ (compared to the significantly faster all-sky survey observing mode). Each portion of the sky is mapped twice (known as a round-trip scan) in order to improve the detection redundancy. Each observation is made up of two round-trip scans, offset by half of the 480-arcsec field of view, in the cross-scan direction. The final map is made up of a mosaic of observations, each offset in the cross-scan direction by 450 arcsec in order to provide a uniformly sampled map. A reset interval of 1 s was selected to maximise the sensitivity, without saturating the detectors at the positions of the brightest sources. For more details, see \citet[][]{2007PASJ...59S.389K}.

The data were reduced using the AKARI official pipeline, version 20070714, developed by the AKARI data-reduction team \citep[][]{akari_fis_manual}. 

\subsection{SCUBA Data}\label{obs_scuba}
The sub-mm data presented in this study were obtained using the Submillimetre Common User Bolometer Array (SCUBA -- \citealp{1999MNRAS.303..659H}) on the JCMT. SCUBA takes observations at 450 and 850~$\mu$m simultaneously through the use of a dichroic beam-splitter. The telescope has a resolution of 8 arcsec at 450$~\mu$m and 14 arcsec at 850~$\mu$m. The data presented here were acquired from the JCMT data archive, operated by the Canadian Astronomy Data Centre.

The observations were carried out over 5 separate nights between April 2002 and December 2003 using the scan-map observing mode. A scan-map is made by scanning the array across the sky, using a scan direction of 15.5$^\circ$ from the axis of the array in order to achieve Nyquist sampling. The array is rastered across the sky to build up a map several arcminutes in extent \citep{SURF}.

Time-dependent variations in the sky emission were removed by chopping the secondary mirror at 7.8 Hz. The size of a scan-map is larger than the chop throw, therefore each source in the map appears as a positive and a negative feature. In order to remove this dual-beam function, each region is mapped six times, using chop throws of 30, 44 and 68 arcsec in both RA and Dec \citep{1995mfsr.conf..309E}. The dual-beam function is removed from each map in Fourier space by dividing each map by the Fourier transform of the dual-beam function, which is a sinusoid. The multiple chop-throws allow for cleaner removal of the dual beam function in Fourier space. The maps are then combined, weighting each map to minimise the noise introduced at the spatial frequencies that correspond to zeroes in the sinusoids. Finally the map is inverse Fourier transformed, at which point it no longer contains the negative sources \citep{SURF}.

The sub-mm zenith opacity at 450 and 850$~\mu$m was determined using the `skydip' method and by comparison with polynomial fits to the 1.3~mm sky opacity data, measured at the Caltech Submillimeter Observatory \citep{2002MNRAS.336....1A}. The sky opacity at 850$~\mu$m varied from 0.18 to 0.37, with a median value of 0.26. These correspond to a 450$~\mu$m opacity range of 0.84 to 2.05, and a median value of 1.36. 

The data were reduced in the normal way using the SCUBA User Reduction Facility, SURF \citep{SURF}. Noisy bolometers were removed by eye, and the baselines, caused by chopping onto sky with a different level of emission, were removed using the {\small MEDIAN} filter. Calibration was performed using observations of the planets Uranus and Mars \citep[][]{1986Icar...67..289O,1993Icar..105..537G}, and the secondary calibrator CRL618 \citep{1994MNRAS.271...75S} taken during each shift. We estimate that the calibration uncertainty is $\pm 5\%$ at 850$~\mu$m and $\pm 15\%$ at 450~$\mu$m, based on the consistency and reproducibility of the calibration from map to map. 

\begin{figure*}
\includegraphics[angle=0,width=120mm]{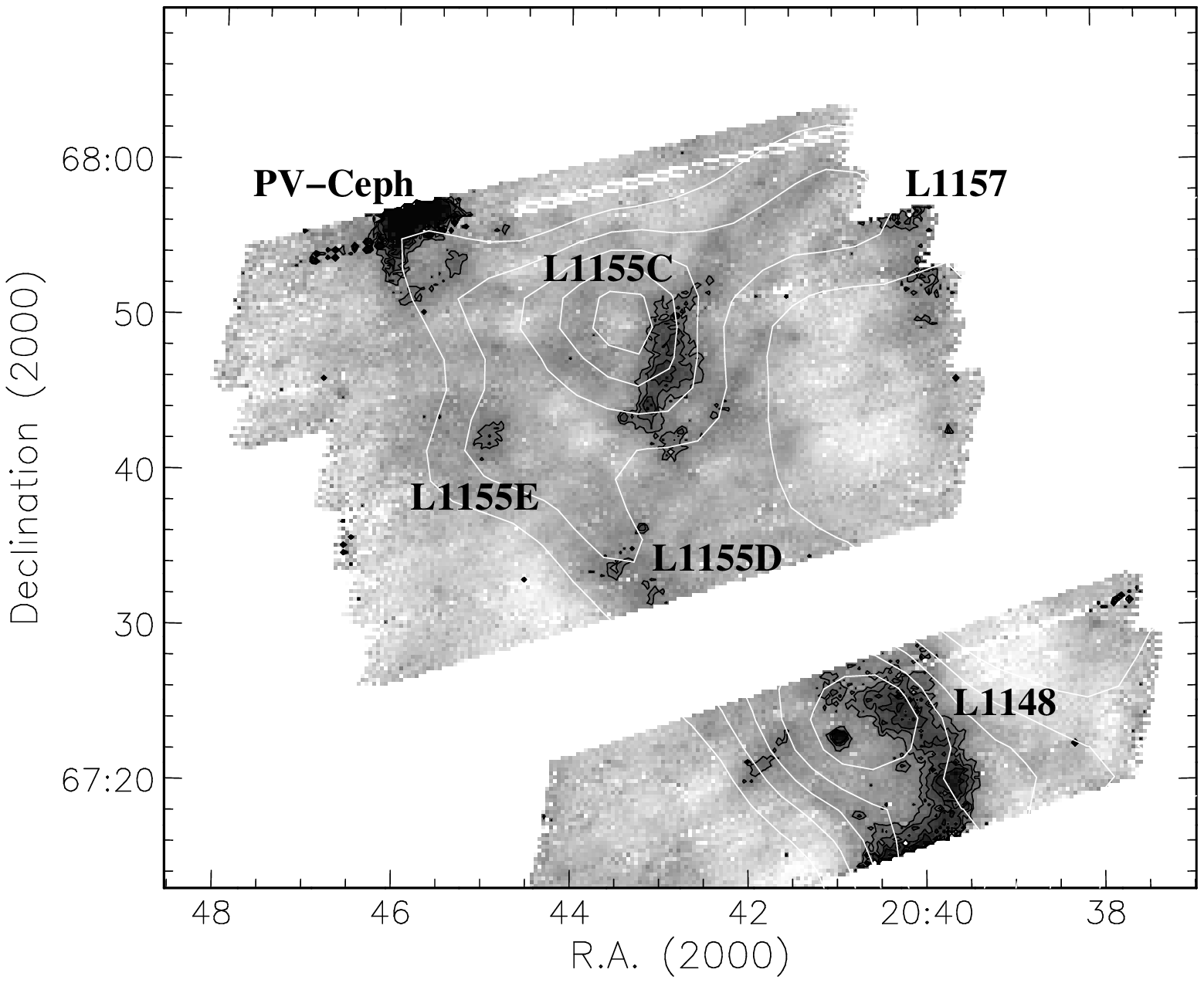} 
\caption{Greyscale image of the 90~$\mu$m Akari WIDE-S data of L1155 and L1148 with black contours overlaid. Contours start at 9.7~MJysr$^{-1}$ with intervals of 0.3~MJysr$^{-1}$. White contours show the \citet{2005PASJ...57S...1D} visual extinction starting at Av=1 with intervals of Av=0.5.  The different clumps have been labeled. Note the offsets between the 90~$\mu$m peaks and the extinction peaks. The bright unresolved object near the extinction peak of L1148 is the very low luminosity Class 0 (VeLLO) L1148-IRS.}
\label{fig_akari90}
\end{figure*}

\begin{figure*}
\includegraphics[angle=0,width=120mm]{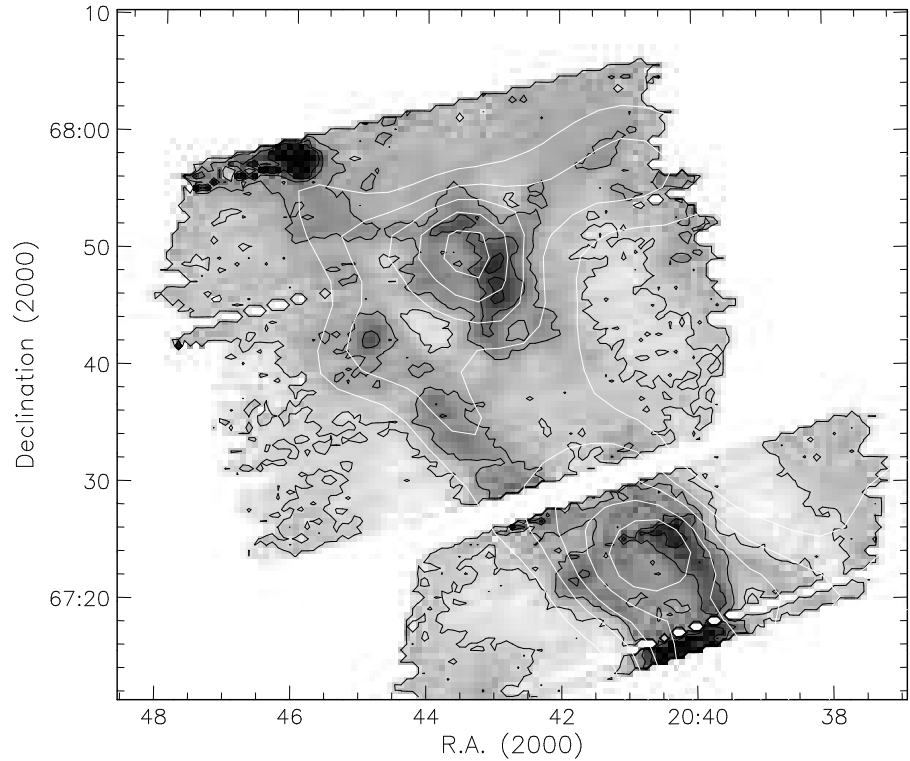}
\caption{Greyscale image of the 140~$\mu$m Akari WIDE-L data of L1155 and L1148 with black contours overlaid. Contours start at 25~MJysr$^{-1}$ with intervals of 5~MJysr$^{-1}$. White contours show the \citet{2005PASJ...57S...1D} visual extinction starting at Av=1 with intervals of Av=0.5. Note again the offsets between the 140~$\mu$m peaks and the extinction peaks.} 
\label{fig_akari140}
\end{figure*}

\begin{figure*}
\includegraphics[angle=0,width=120mm]{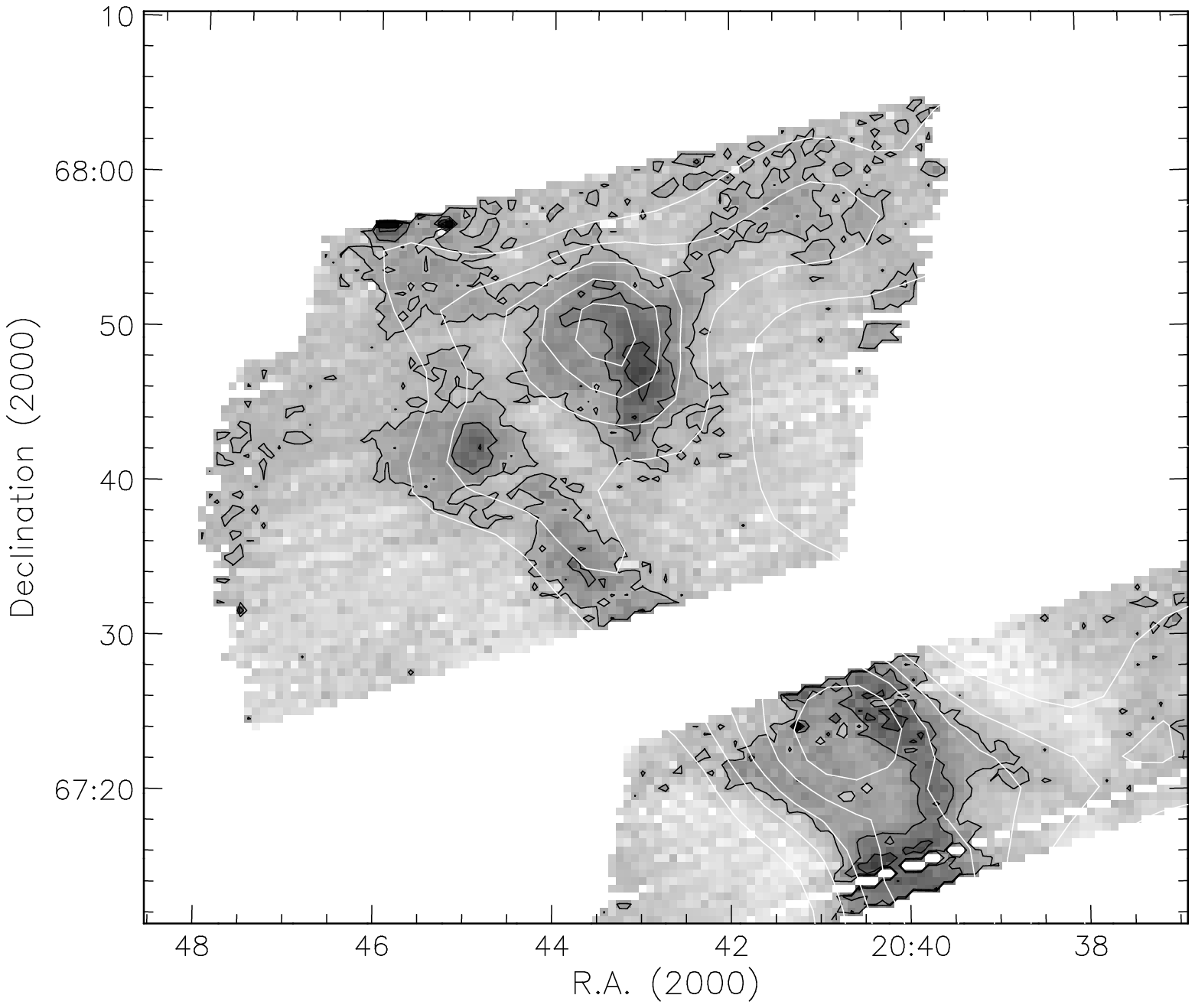}
\caption{Greyscale image of the 160~$\mu$m Akari data of L1155 and L1148 with black contours overlaid. Contours start at 25~MJysr$^{-1}$ with intervals of 5~MJysr$^{-1}$. White contours show the \citet{2005PASJ...57S...1D} visual extinction starting at Av=1 with intervals of Av=0.5. Once again the far-ir emission is offset from the extinction peaks.} 
\label{fig_akari160}
\end{figure*}

\subsection{ISO Data}\label{obs_ISO}
The Infrared Space Observatory (ISO) satellite was a mid- and far-infrared telescope that was launched by the European Space Agency (ESA) in 1995 and operated until 1998 \citep[][]{1996A&A...315L..27K}. The imaging photo-polarimeter \citep[ISOPHOT--][]{ISOPHOT} was used in its over-sampled mapping mode (PHT32) to observe a number of pre-stellar cores, including L1155C in Cepheus \citep[][]{2002MNRAS.329..257W}. Data were obtained at 90, 170 and 200~$\mu$m for this core with diffraction-limited resolutions of 40 -- 85 arcsec. See \citet[][]{2002MNRAS.329..257W} for more details regarding the observations and reduction of these data. The absolute calibration errors were estimated to be 30\% \citep[][]{2002MNRAS.329..257W}.

The characteristics of each of the filters used in Akari \citep[][]{akari_fis_manual}, ISOPHOT \citep[][]{ISOPHOT} and SCUBA \citep[][]{1999MNRAS.303..659H} are given in Table~\ref{table_filters}.


\section{Results}\label{results}

\subsection{Akari}

Figures~\ref{fig_akari90}, \ref{fig_akari140} and \ref{fig_akari160} show the L1155/L1148 region at 90, 140 and 160~$\mu$m respectively. White contours showing the location of the optical extinction \citep[][]{2005PASJ...57S...1D} are overlaid on to each map. These contours are plotted to highlight any differences in the morphology of the Akari data at the three different wavelengths. The only source detected in the N60 data is the point source L1148-IRS. These data are therefore not shown here.

There are a number of different sources seen in the data. These sources are labelled in Figure \ref{fig_akari90}. L1155 \citep{1962ApJS....7....1L} is located in the centre of the northern portion of the map, close to the northern extinction peak (Lynds clouds were discovered as opaque regions in the sky, therefore they would be expected to correlate with extinction peaks). L1148 is located in the southern portion of the map, and is also seen as an extinction peak. 

Since the study of extinction peaks in optical maps by \citet[][]{1962ApJS....7....1L}, the L1155 clump has been subdivided using CO and NH$_3$ observations \citep[e.g.][]{1983ApJ...264..517M}, and a number of these sub-clumps are seen in the Akari data. The extinction peak is centred on L1155C. L1155E is located 10 arcmin east of L1155C, and L1155D is approximately 12 arcmin southeast of L1155C. These have been labelled on Figure~\ref{fig_akari90}. 

We believe that L1155C is a pre-stellar core due to the detection of a centrally-condensed object at submillimetre wavelengths with a high density contrast with its surroundings (Section \ref{scuba_results} -- see also \citealp{2007prpl.conf...33W}). We also class this as isolated due to the large distance to the nearby cores, compared to more clustered star-forming regions such as Ophiuchus and Orion \citep[e.g.][]{1998A&A...336..150M,2007MNRAS.374.1413N,2008MNRAS.391..205S}.

The other fragments of L1155 may also be pre-stellar in nature, or they may simply be transient starless cores. We have insufficient data to reliably determine their status, and so do not consider them in this paper. 

The emission at the far western edge of the map belongs to L1157, which contains the Class 0 protostar IRAS 20386+6751 \citep[][]{1995ApJ...443L..37T,2007ApJ...670L.131L}, the driving source of a powerful bipolar outflow. The bright source seen on the north-eastern edge of the maps is the well-studied Herbig Ae/Be star PV Ceph \citep[][]{1994ApJ...433..199L}, which is the source of the parsec scale outflow generating the HH 315 knots \citep[][]{2002ApJ...575..911A}. PV Ceph is notable for having an unusually high velocity of over 20 kms$^{-1}$ \citep[][]{2004ApJ...608..831G}, and is believed to have been ejected from the nearby star-forming cluster NGC 7023.

L1148 contains the very low luminosity Class 0 (VeLLO) L1148-IRS \citep{2005AN....326..878K}. This is a very low-mass Class 0 protostar, and is believed to be either the precursor of a brown dwarf or a low-mass star, depending on the amount of gas it accretes in the future. This source is seen in both the N60 and the WIDE-S data.

In Figures \ref{fig_akari90}, \ref{fig_akari140} and \ref{fig_akari160} we clearly see that the far-infrared emission peak of L1155C is not coincident with the extinction peak. Instead, we see the far-infrared emission curving around the south-west edge of the extinction peak. The shortest wavelength shows this effect most prominently, peaking furthest away from the maximum extinction. The same is true in the L1148 cloud, where the far-infrared emission wraps around the western edge of the extinction. The extinction peaks of the other sub-clumps of L1155 are unresolved from the larger L1155C and L1148 peaks, therefore we cannot detect any offsets in these clouds. 

The far-infrared emission detected by Akari traces the density of the emitting dust weighted by its temperature, whereas the extinction map is based solely on the column density structure of the material providing the extinction. We therefore believe that the offset between the far-infrared and the extinction peaks can be explained as a temperature gradient across the core. This is considered in more detail in Section~\ref{analysis}.

\subsection{SCUBA}\label{scuba_results}

\begin{figure}
\includegraphics[angle=0,width=70mm]{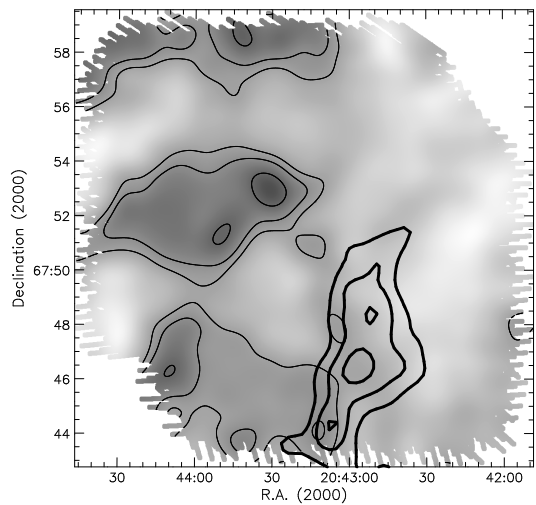} 
\caption{The SCUBA map of L1155C at 850~$\mu$m with thin black contours overlaid. The data have been smoothed to a FWHM of 60 arcsec to improve the signal to noise ratio and to highlight the larger scale structures in the data. The noise level in the central part of the smoothed map is 3~mJy/60-arcsec-beam. The thin contour levels are 3, 6 and 12-$\sigma$. We note that the far north and south portions of the map have lower integration times, therefore the noise-level is higher in these regions. The structure seen in these parts of the map are therefore less significant than in the map centre. The thick contours show the 90~$\mu$m Akari WIDE-S data, also smoothed to 60 arcsec for comparison, with contour levels as in Figure \ref{fig_akari90}.}
\label{scuba_data}
\end{figure}

The 850~$\mu$m SCUBA data of L1155C are shown in Figure~\ref{scuba_data}. Contours showing the 90~$\mu$m Akari WIDE-S data are included to highlight the position of the SCUBA core compared to the location of the core seen at other wavelengths. The sub-mm data shown have been smoothed from 14~arcsec to 60~arcsec. This was done in order to increase the signal to noise ratio of the data and to highlight the larger scale structures for comparison with the cloud models described in Section \ref{rt-1155}.

We note that there is a small offset between the peak of the SCUBA data, and the peak of the \citet{2005PASJ...57S...1D} extinction map. However, this offset is less than the 6 arcmin resolution of the extinction map, therefore we don't believe that this offset is significant. We also note that the morphology of the SCUBA data shows a very close agreement with a higher resolution extinction map generated from data taken with the IRAC camera on board the Spitzer space telescope (Chapman, 2009, priv. comm.). We therefore believe that the 850~$\mu$m peak represents the location of the maximum column density of L1155.

None of the sources in the mapped area were detected at 450~$\mu$m, therefore these data have not been shown here. The reason for the non-detection of any sources is probably due to the poor atmospheric transmission at this wavelength. These data are used to provide upper-limits at this wavelength in Section~\ref{analysis}.

\begin{table}
\caption{The characteristics of the different wavebands and the 1-$\sigma$ rms in each of the maps. $^\dag$ISOPHOT rms values taken from \citet{2002MNRAS.329..257W}}
\label{table_filters}
\begin{tabular}{llccc}\hline
Instrument 	& Filter	& Central 	& Bandwidth	& rms	\\
		& name		& wavelength	& ($\mu$m)	& (MJy/sr)\\
		& 		& ($\mu$m)	& 		& 		\\ \hline
Akari		& N60		& 65		& 21.7		& 0.5~		\\		
~~~~"		& WIDE-S	& 90		& 37.9		& 0.1~		\\
~~~~"		& WIDE-L	& 140		& 52.4		& 0.7~		\\
~~~~"		& N160		& 160		& 34.1		& 1.3~		\\ 
ISOPHOT		& C-90		& 90		& 40		& 0.7$^\dag$	\\
~~~~"		& C-160		& 170		& 50		& 5.7$^\dag$	\\
~~~~"		& C-200		& 200		& 30		& 6.5$^\dag$	\\
SCUBA		& 450		& 450		& 66		& 120~		\\
~~~~"		& 850		& 850		& 22		& 2.7~		\\ \hline
\end{tabular}
\end{table}

\subsection{ISO}

\begin{figure*}
\includegraphics[angle=0,width=175mm]{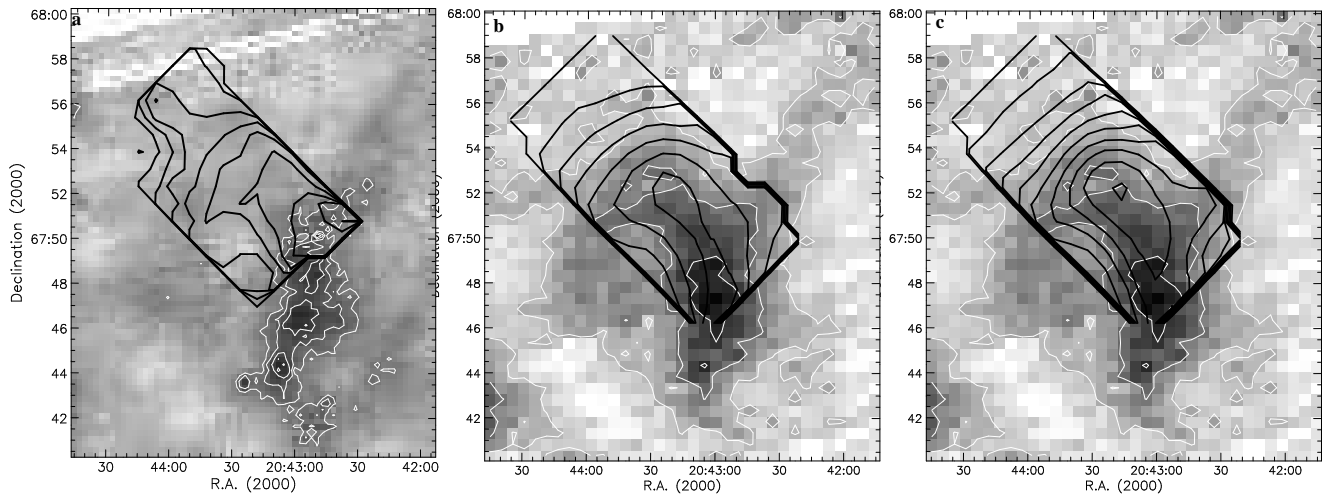} 
\caption{The ISOPHOT maps of L1155C at (a) 90~$\mu$m, (b) 170~$\mu$m and (c) 200~$\mu$m, shown as thick black contours. The contour levels are (a) 18~MJySr$^{-1}$, with 0.5~MJySr$^{-1}$ intervals, (b) 40~MJySr$^{-1}$, with 10~MJySr$^{-1}$ intervals, and (c) 30~MJySr$^{-1}$, with 10~MJySr$^{-1}$ intervals. The background greyscale and white contours show a close-up of the Akari maps at (a) 90~$\mu$m and (b) \& (c) 160~$\mu$m, with contour levels as in Figures~\ref{fig_akari90} and~\ref{fig_akari160}. The similarity between the ISO data and the Akari data is striking.}
\label{iso_data}
\end{figure*}

Figure~\ref{iso_data} shows the 90, 170 and 200~$\mu$m ISOPHOT data of L1155C from \citet[][]{2002MNRAS.329..257W}, displayed as contours overlaid on close-ups of the Akari 90 and 160~$\mu$m maps. The 170~$\mu$m ISOPHOT data are consistent with the 160~$\mu$m Akari data, showing a ridge of emission that stretches from the 160~$\mu$m Akari peak northwards. The 200~$\mu$m ISOPHOT data peak at the northern extent of this ridge. The 90~$\mu$m ISOPHOT data miss most of the emission seen in the 90~$\mu$m Akari data. The reason for this is two-fold. The 90~$\mu$m emission lies further south than the longer wavelength emission, and also the location of the detectors on the ISOPHOT focal plane meant that the 90~$\mu$m detector was offset, and in this case was unfortunately shifted slightly to the northeast. However, the emission from the edge of the L1155C core detected by ISOPHOT at 90~$\mu$m is consistent with the Akari map at this wavelength. 

The ISOPHOT data therefore show the same trend for shorter wavelength emission to peak further away from the extinction peak, and back up the hypothesis of a temperature gradient. The rms noise levels in each of the Akari, ISO and SCUBA maps is given in Table~\ref{table_filters}.

\section{Analysis}\label{analysis}

\subsection{Temperature Gradient}\label{temp}

\begin{figure}
\includegraphics[angle=270,width=87mm]{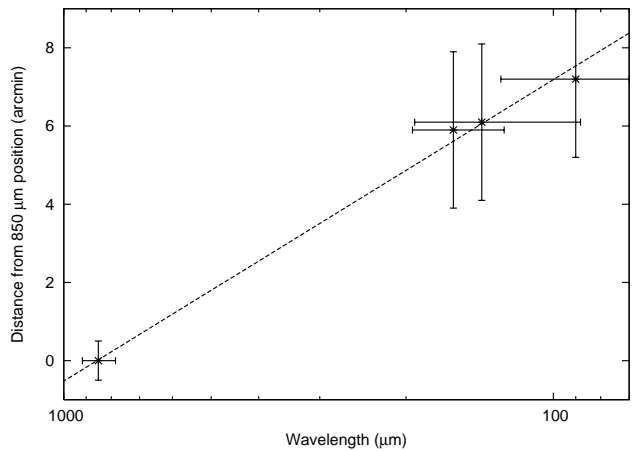} 
\caption{A graph showing the position of the peak of emission of L1155C as a function of wavelength, based on the Akari and SCUBA data. The position measured is at the closest point to a straight line joining the 90 and 850~$\mu$m peaks. The dashed line shows a fit to the data. The horizontal error-bars show the filter bandwidth. The vertical error-bars show the estimated positional accuracy in determining the peak of emission.}
\label{position_graph}
\end{figure}

In the previous section we noted that the position of the emission peak of L1155C depends on the wavelength being observed, with the shorter wavelength radiation peaking furthest from the centre of the core. This trend is highlighted in Figure~\ref{position_graph}, which shows the position of the emission-peak as a function of wavelength. This shows that there is a monotonic shift of the peak with wavelength. 

As stated in the previous section, we believe that this is caused by a temperature gradient across the core. \citet[][]{Kirk_spitzer} have shown that there are no embedded sources in L1155C, therefore the most likely cause of a temperature gradient is a heating source (presumably a star) located outside L1155C. We discuss the possible identity of this heating source in Section~\ref{heating_source}.

\begin{figure*}
\begin{minipage}{160mm}
\includegraphics[angle=0,width=80mm]{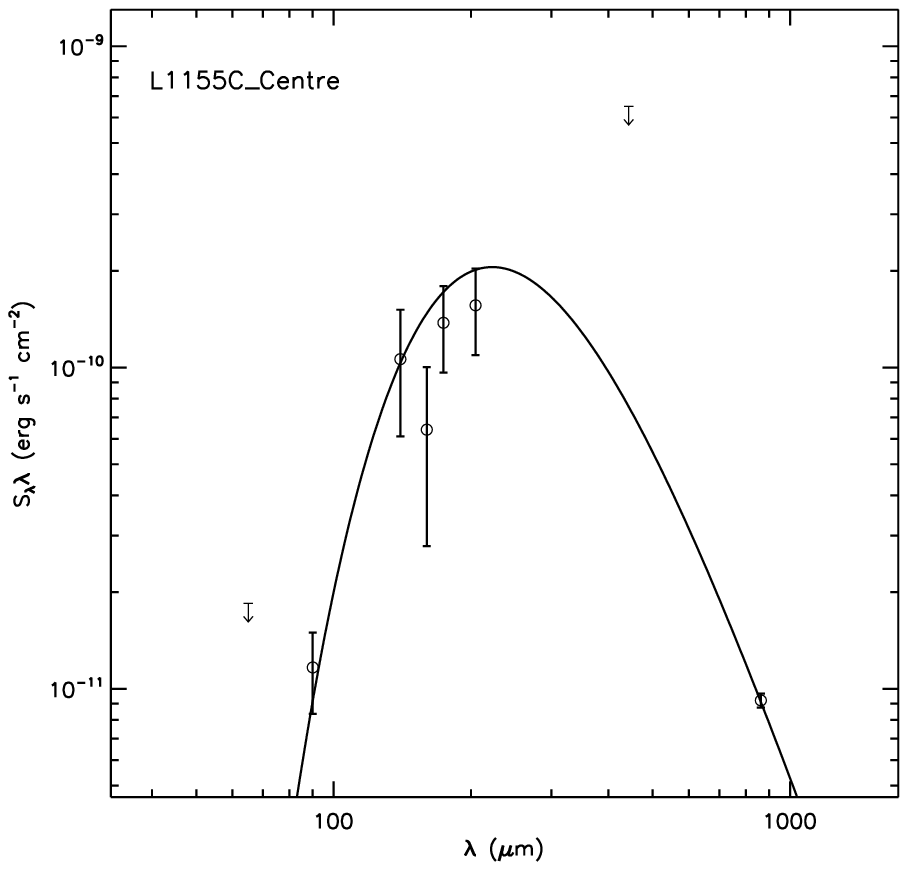} 
\includegraphics[angle=0,width=80mm]{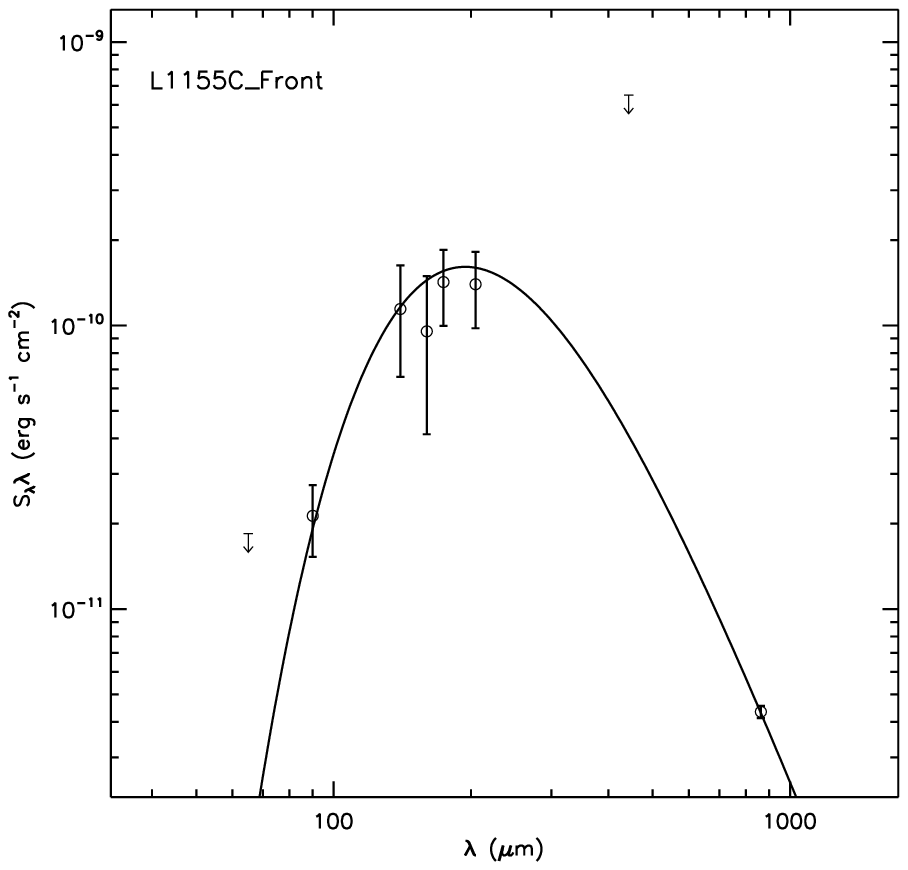}
\end{minipage}
\caption{SEDs for the centre and front positions. In both cases, the background flux density was removed from all points. The plotted line is a modified blackbody fit to the data. The error-bars show the calibration uncertainty at each wavelength (see Section \ref{observations}). 3-$\sigma$ upper-limits are given for the Akari 65~$\mu$m and SCUBA 450~$\mu$m data. The best-fit temperatures are 11.9~K and 13.6~K for the centre and front positions respectively. See text for the parameters of each fit and for further details.} 
\label{seds}
\end{figure*}

To investigate the temperature gradient hypothesis, we measured the flux densities at each of the wavelengths\footnote{The 90~$\mu$m ISOPHOT data were not included in this analysis because the data do not cover the positions of both the 850~$\mu$m SCUBA peak, and the 90~$\mu$m Akari peak. See Section~\ref{results} for more details.} at two different positions. The first position is the centre of L1155C. To determine this position, we use the peak in the column density, which is the most likely the position of the volume density peak. We measure the position of the column density peak from the 850~$\mu$m SCUBA map, because this wavelength is the least affected by the dust temperature, and is hence the most reliable tracer of column density.  We note that there is material surrounding the peak, which is detected at 850~$\mu$m, and this material appears to contain sub-fragments. However, we do not include this material in our estimate of the position of the centre of L1155C, as we do not know what will happen to this material in the future. It could accrete onto the main L1155C core, fragment and form other stars, or it may simply disperse.  

As stated above, the morphology of the 850~$\mu$m emission shows excellent agreement with a higher resolution extinction map, generated with data from the Spitzer space telescope. Therefore we don't consider the offset between the 850~$\mu$m peak and the lower resolution \citet{2005PASJ...57S...1D} to be significant. The centre position is therefore taken to be 20:43:30, +67:52:51 (J2000).

The second position (20:42:58, +67:46:27, J2000) is the peak of the 90~$\mu$m Akari map, which we hereafter refer to as the front of L1155C because if our hypothesis is correct, this region is the closest to the external heating source. 
The front position is located 0.5~pc away from the centre position.
The flux densities at these positions were measured in circular apertures with a diameter of 2 arcminutes, and are given in Table~\ref{aperture_fluxes}. 

The JCMT uses a chopping secondary mirror to remove the atmospheric emission (see Section~\ref{obs_scuba}). A consequence of this is that SCUBA does not measure the total power in the map, but only the power relative to the rest of the map. Akari and ISO, being satellites, have no need to chop and so they do measure the total power at each point on the sky. In order to compare the flux densities measured using ISO and Akari with those measured using SCUBA, all the fluxes were made relative to a common point, as far away from L1155C as possible. 

\begin{figure*}
\includegraphics[angle=0,width=150mm]{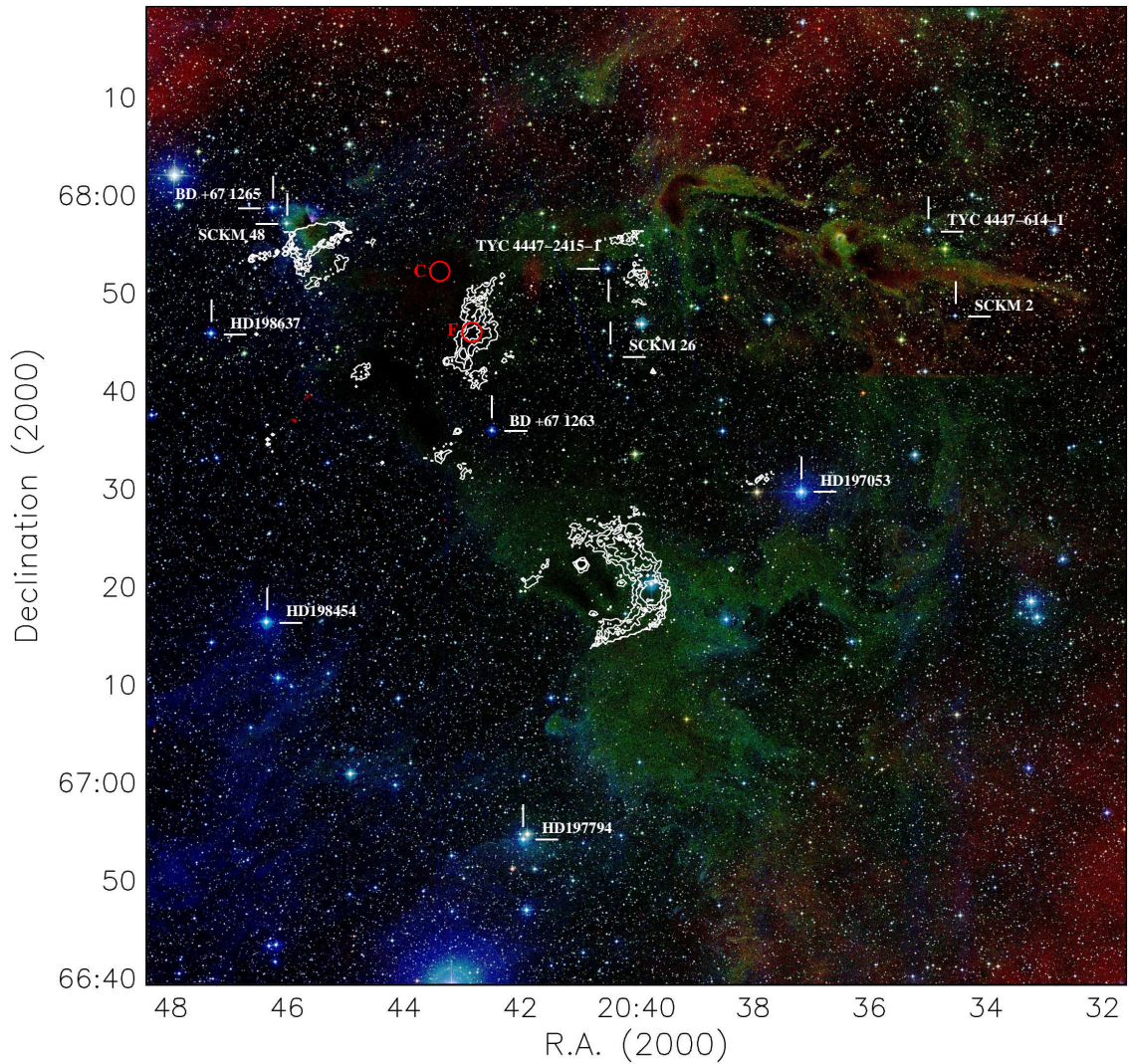} 
\caption{A colour composite of the region around L1155 taken from the Digital Sky Survey (DSS). The red, green \& blue channels show data taken using the infrared (850~nm), red (650~nm) and blue (480~nm) filters respectively. The white contours show the location of the Akari 90~$\mu$m emission (as for Figure~\ref{fig_akari90}). The positions of the candidate heating sources are labeled (see Table \ref{stars} for details). The front and centre positions are also marked with red circles.}
\label{colour_composite}
\end{figure*}

Spectral energy distributions (SEDs) were plotted for both the centre and the front positions. These SEDs are shown in Figure~\ref{seds}. Upper limits are shown for the 65~$\mu$m Akari data and the 450~$\mu$m SCUBA data, where no signal was detected.   

We fit a modified blackbody function to the data at each of the two positions in the normal manner \citep[see, e.g.][]{2002MNRAS.329..257W}. The best fits to the data are found to have a temperature of 11.9~K for the centre position, and 13.6~K for the front position. The value of $\beta$ (the dust emissivity index) used to produce these fits was 1.5. This differs from the canonical value of 2.0, which gave a poorer fit. However, we note that all values of $\beta$ gave the same temperature difference of $\sim 2$~K when constrained to be the same at both positions. Following \citet{2007MNRAS.375..843K} the critical wavelength ($\lambda_c$) was set equal to 50~$\mu$m \citep[see][]{1983QJRAS..24..267H}. Figure~\ref{seds} indicates that the calibration may be marginally too low for the 160~$\mu$m N160 Akari data, as the flux density at this wavelength is systematically lower than the neighbouring wavelengths.

These data are therefore in agreement with our hypothesis that the front of the L1155C core is warmer than the centre. In the following section we consider the possible heating sources that could be the cause of this temperature gradient.

\begin{table}
\caption{Background-subtracted flux densities measured at the centre (20:43:30, +67:52:51, J2000) and front (20:42:58, +67:46:27, J2000) positions in the Akari, ISOPHOT and SCUBA maps. Flux densities are measured in a 2 arcmin diameter aperture. 3-$\sigma$ upper-limits are given for the Akari 65~$\mu$m and SCUBA 450~$\mu$m data.}
\label{aperture_fluxes}
\begin{tabular}{lcrr}\hline
Instrument 	& Wavelength	& \multicolumn{2}{c}{Flux density (Jy)}		\\
		& ($\mu$m)	& Centre	& Front		\\ \hline
Akari		&  ~65		& $<$0.4	& $<$0.4	\\
~~~~"		&  ~90		& 0.4		& 0.6		\\
~~~~"		&  140		& 5.0		& 5.3		\\
~~~~"		&  160		& 3.4		& 5.1		\\ 
ISOPHOT		&  170		& 8.0		& 8.3	  	\\
~~~~"		&  200		& 10.7		& 9.5 		\\
SCUBA		&  450		& $<$96		& $<$96		\\
~~~~"		&  850		& 2.7		& 1.3 		\\ \hline
\end{tabular}
\end{table}

\subsection{Heating Sources}\label{heating_source}

The SIMBAD database was searched for all stars within 5 degrees of L1155. This corresponds to 28 pc at the canonical distance to the Cepheus molecular cloud of 325 pc. We only considered stars which were approximately the same distance away from the sun as the Cepheus molecular cloud (see Section~\ref{intro}), though we note that the uncertainty in the distances both to the cloud and to individual stars is large. Only stars of spectral type O, B, A and F were considered, as later-type stars would not be luminous enough to have a significant heating effect. The radiant flux due to each star at the position of the 90~$\mu$m intensity peak was calculated to determine the most likely candidates. Table~\ref{stars} gives the details of these candidates. 

The flux due to each star at the position of L1155C depends crucially on the distance between the star and the cloud. As this distance is not known accurately, upper and lower limits were estimated. The lower limit makes the assumption that the star and the cloud lie at the same distance from the sun, i.e. the sun-cloud-star makes an angle of 90$^{\circ}$. It is assumed that this distance is the measured distance to the star, rather than the indirectly measured distance to the cloud. For the upper limit, we assumed that the sun-cloud-star makes an angle of 45$^\circ$ or 135$^\circ$, therefore the distance between the star and the cloud is $\sqrt{2}$ times larger than the distance in the plane of the sky. This assumption is based on the observed geometry of L1155C at the shorter wavelengths. The core appears to being heated from one side, rather than the front or the back.

Figure~\ref{colour_composite} shows a colour composite of the digitized sky survey (DSS, \citealp{1994IAUS..161..167L}), obtained using the SkyView interface \citep{1994ASPC...61...34M}. The red, green and blue channels represent the infrared, red and blue filters respectively. The contours of the 90~$\mu$m Akari data from Figure~\ref{fig_akari90} are shown, to indicate the location of L1155C. The positions of the most likely heating sources are also highlighted. 

Table~\ref{stars} indicates that the most likely heating source is the nearby A6V star BD+67 1263, as it produces an order of magnitude more heating flux that any other star. In the following section we use a radiative transfer simulation to determine if this or any other of the stars in Table~\ref{stars} is appropriate for causing the temperature gradient that we see in the data.

\begin{table*}
\caption{Details of each of the possible heating sources in the vicinity of L1155. Column 2 shows the distance of each star from the sun. Columns 5 \& 6 give the minimum and maximum distance between each star and L1155C (see text for details). Columns 7 \& 8 give the corresponding flux due to the star at the front of L1155C. Distance references:  
1. \citet[][]{1992BaltA...1..149S}; 2. \citet[][]{1997A&A...323L..49P};}
\label{stars}
\begin{tabular}{lllclcccrr}\hline
Star		& ~~R.A.	& ~~Dec.		& Distance & Spectral	& L 		& \multicolumn{2}{c}{Distance from front}	 & \multicolumn{2}{c}{Flux at front of L1155C} \\
		& ~(2000)	& ~(2000) 	& (pc)	   & ~~type	& (L$_\odot$)	& \multicolumn{2}{c}{of L1155C (pc)}		 & \multicolumn{2}{c}{({\rm $10^{-9}Wm^{-2}$})} \\ 
		&		&		&	   &		&	   	& Min	& Max				 	& Upper		& Lower			\\ \hline
BD+67 1263	& 20:42:35.7	& 67:36:35.1	& 280$^1$  & ~~A6V 	& 22	   	& 0.7	& 1.0				 	& 1024		& 521			\\
HD197053	& 20:37:04.1	& 67:30:15.5	& 246$^2$  & ~~B9V	& 42	   	& 2.7	& 3.8				 	& 148		& 74			\\
SCKM 26		& 20:40:28	& 67:44:24	& 480$^1$  & ~~A5IV	& 20		& 1.8	& 2.6					& 148		& 74			\\
BD+67 1265	& 20:46:37.8	& 67:58:58.4	& 340$^1$  & ~~A1V	& 24		& 2.4	& 3.4					& 108		& 54			\\
HD198454	& 20:46:32.2	& 67:16:32.5	& 344$^2$  & ~~B9	& 51		& 3.6	& 5.1					& 98		& 49			\\
TYC 4447-2415-1 & 20:40:31.2	& 67:53:15.2	& 200$^1$  & ~~F6V	& 2		& 0.8	& 1.2					& 81		& 41			\\
SCKM 48		& 20:46:23	& 67:57:24	& 300$^1$  & ~~A1III	& 12		& 1.9	& 2.7					& 78		& 39			\\
HD198637	& 20:47:40.2	& 67:45:55.6	& 330$^1$  & ~~A3V	& 19		& 2.6	& 3.7					& 72		& 36			\\
HD197794	& 20:41:58.3	& 66:54:43.7	& 190$^1$  & ~~A8IV	& 12		& 2.9	& 4.1					& 36		& 18			\\
TYC 4447-614-1	& 20:34:42.1	& 67:56:54.3	& 400$^1$  & ~~A3IV	& 31		& 5.5	& 7.7					& 26		& 13			\\
SCKM 2		& 20:34:15	& 67:47:54	& 340$^1$  & ~~A5V	& 11		& 4.8	& 6.9					& 11		& 6			\\ 
HD194297	& 20:20:42.3	& 66:41:37.4	& 360$^2$  & ~~~~--	& 49		& 15.0	& 21.2					& 5		& 3			\\ \hline
\end{tabular}
\end{table*}

\subsection{Radiative Transfer Modelling - L1155C}\label{rt-1155}

The radiative transfer calculations were performed using the {\sc PHAETHON} code, which is a 3D Monte Carlo radiative transfer code developed by \citet{2003A&A...407..941S}. {\sc PHAETHON} simulates the heating of a cloud by injecting a large number of monochromatic luminosity packets ($L$-packets). These $L$-packets interact (by being absorbed, re-emitted or scattered) stochastically with the cloud. The injection point(s) of the $L$-packets represents the location of the radiation source(s) in the system. 

If an $L$-packet is absorbed, its energy raises the local temperature of the absorbing region. To ensure radiative equilibrium, the $L$-packet is re-emitted immediately with a new frequency chosen from the difference between the local cell emissivity before and after the absorption of the packet \citep{2001ApJ...554..615B,2005NewA...10..523B}. For more details, see \citet{2003A&A...407..941S}. The model can produce images of the core at any wavelength, convolved to any resolution.

The core is illuminated by a combination of the interstellar radiation field (ISRF), and the energy from the nearby heating source (see Table~\ref{stars}). The ISRF is isotropic, whereas the energy from the nearby star is directional. For the ISRF we adopt a revised version of the \citet{1994ASPC...58..355B} interstellar radiation field (BISRF). The BISRF consists of radiation from giant stars and dwarfs, thermal emission from dust grains, cosmic background radiation, and mid-infrared emission from transiently heated small PAH grains \citep{2003cdsf.conf..127A}. The heating from the nearby star is parameterised by the star's radius, surface temperature and its distance from the core.

\begin{figure}
\includegraphics[angle=270,width=87mm]{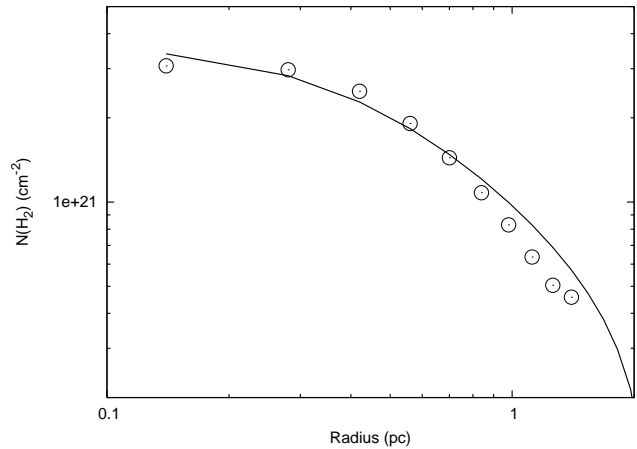} 
\caption{The azimuthally averaged radial column-density profile of L1155C, determined from the \citet{2005PASJ...57S...1D} extinction data (open circles), with the best-fit Plummer-like sphere plotted as a solid line. See text for details. }
\label{density_prof}
\end{figure}               

\begin{figure}
\includegraphics[angle=0,width=87mm]{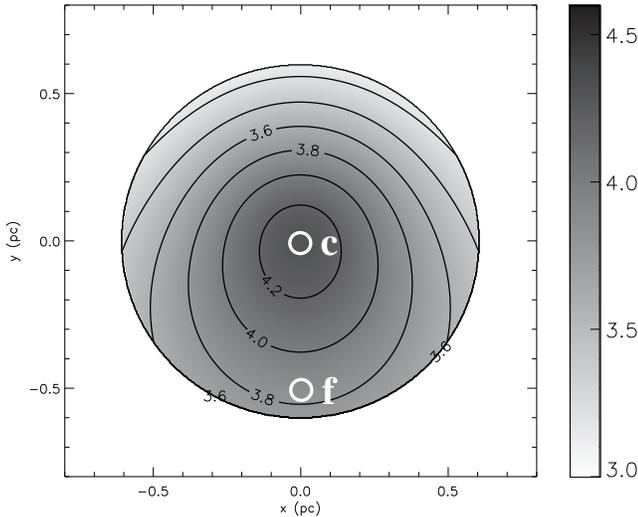} 
\caption{The density profile of the best-fit model of L1155C. The scale on the right shows the logarithm of the volume number density, measured in units of cm$^{-3}$. The heating source is off the bottom of the figure. The centre and front positions are marked with a `c' and an `f' respectively.}
\label{density_map}
\end{figure}

\begin{figure*}
\begin{minipage}{83mm}
\hspace{5mm}
\includegraphics[angle=0,width=68.8mm]{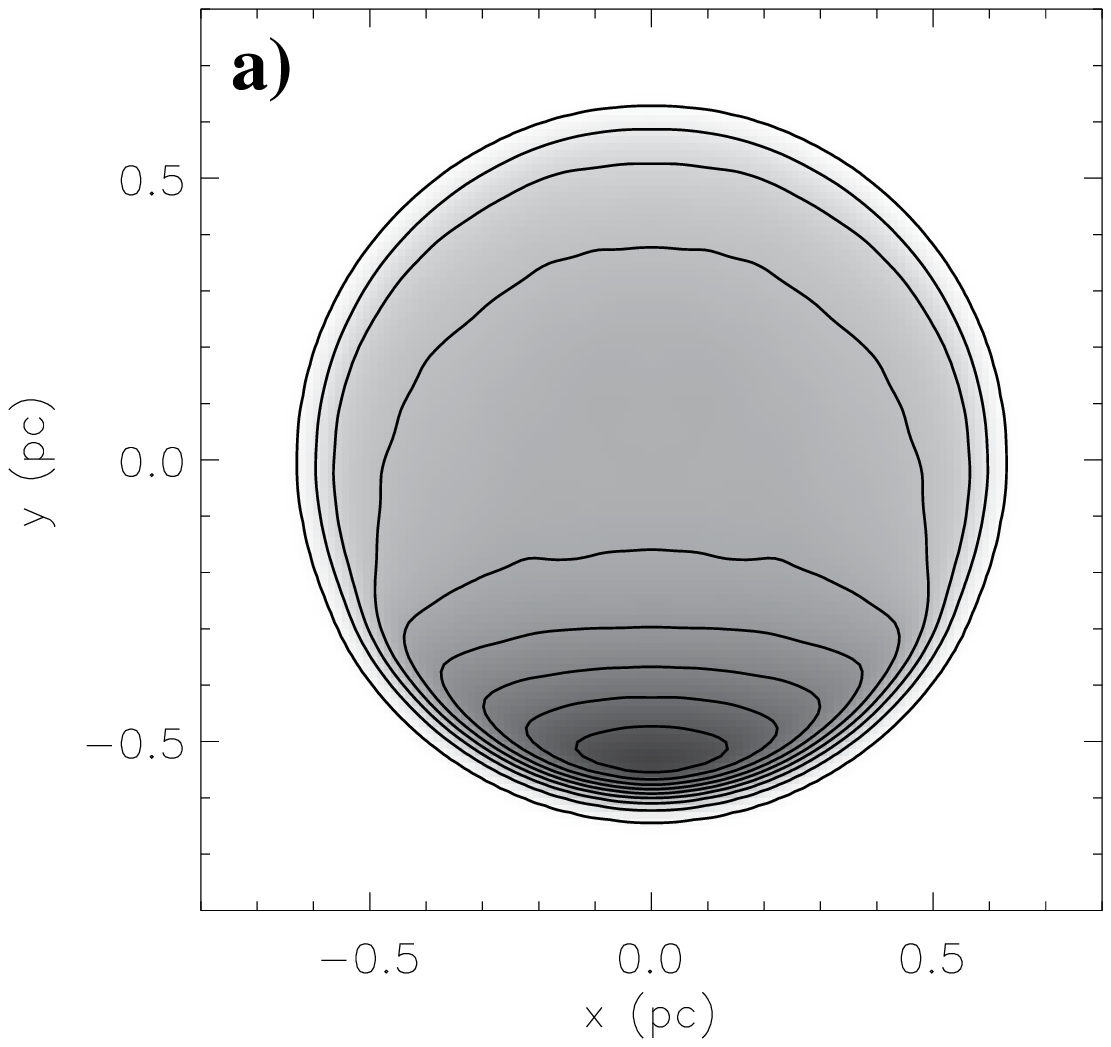} 
\end{minipage}
\begin{minipage}{83mm}
\hspace{-5mm}
\includegraphics[angle=0,width=68.8mm]{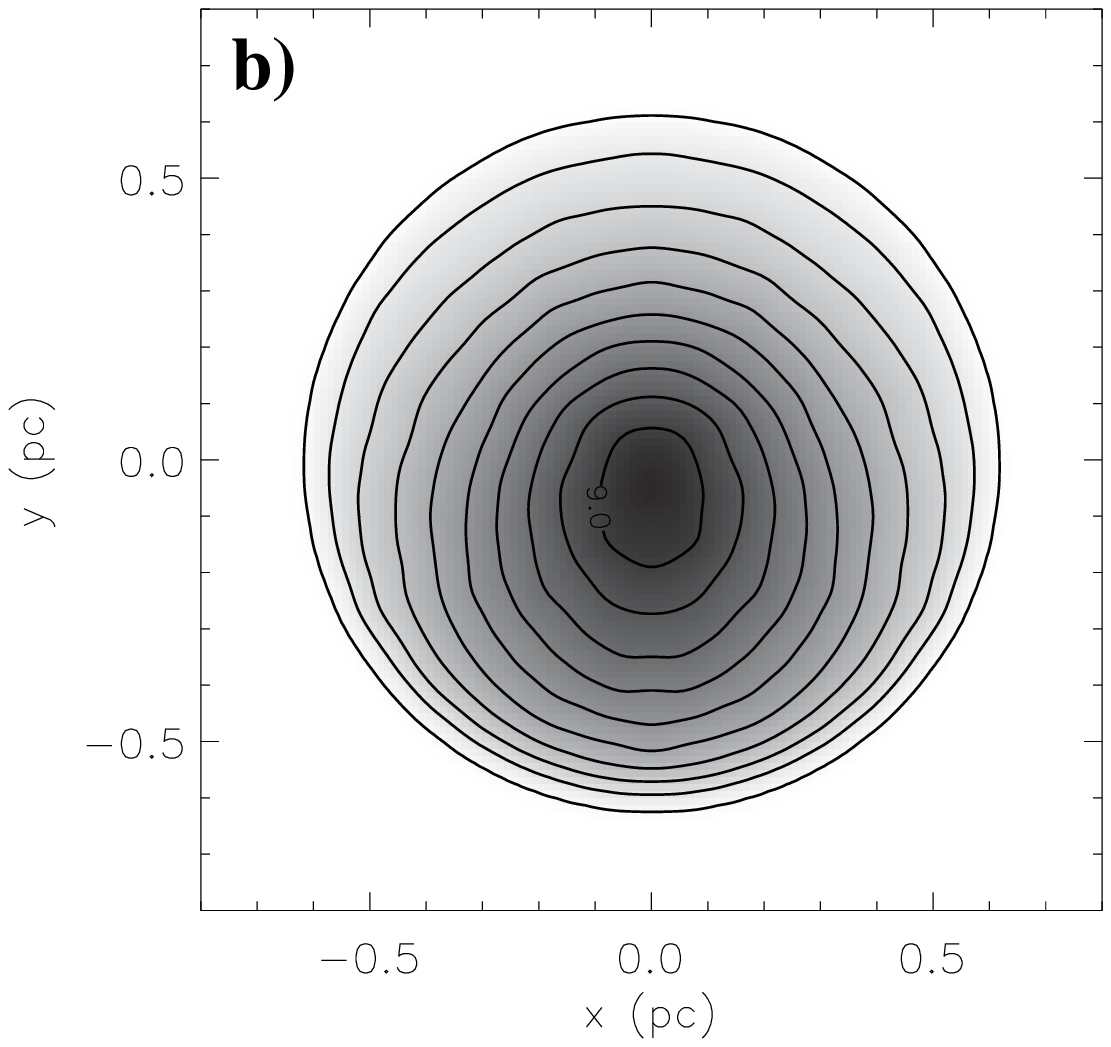}
\end{minipage}
\begin{minipage}{83mm}
\hspace{8.5mm}
\includegraphics[angle=0,width=64.8mm]{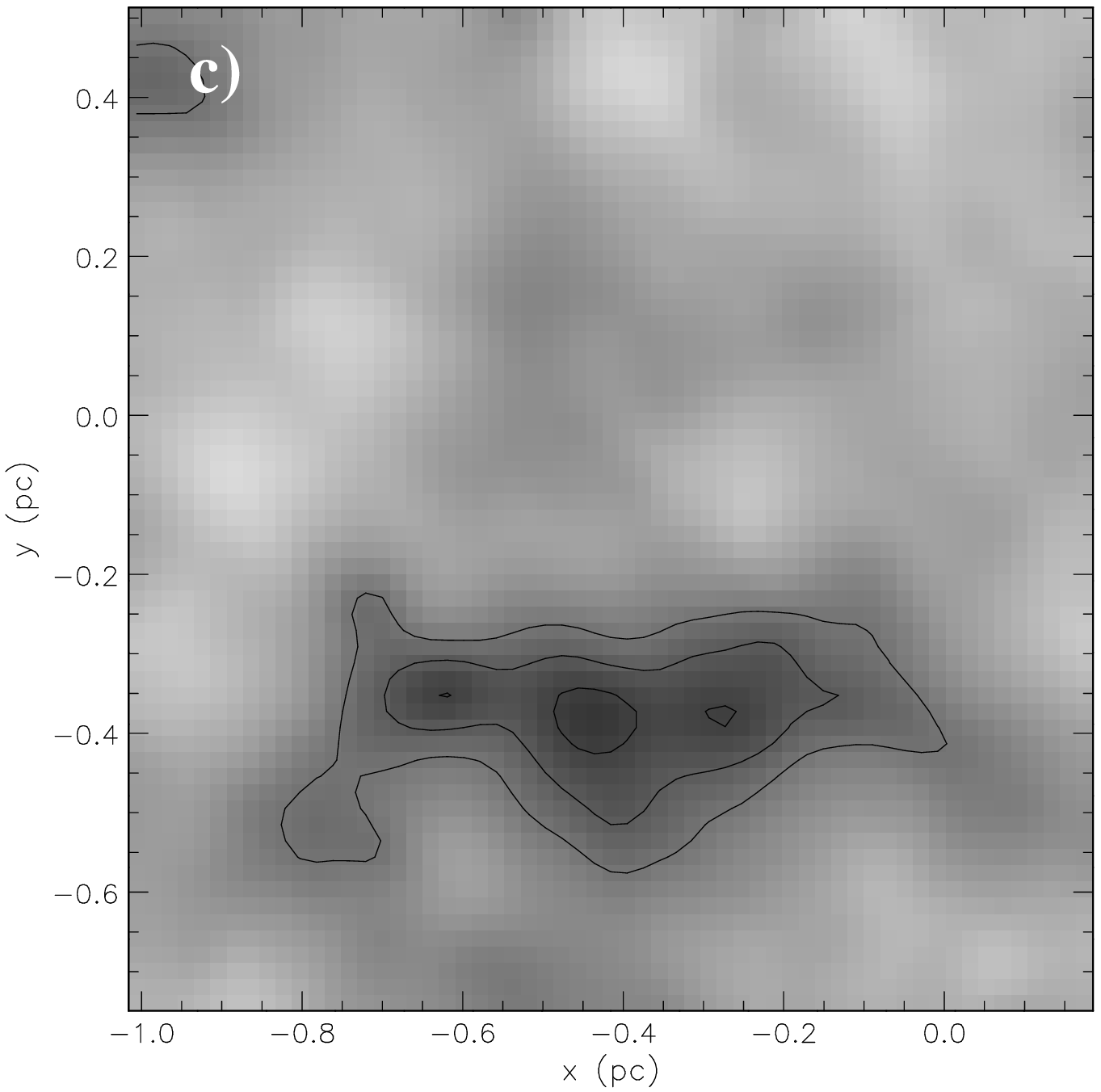} 
\end{minipage}
\begin{minipage}{83mm}
\hspace{-1.5mm}
\includegraphics[angle=0,width=65mm]{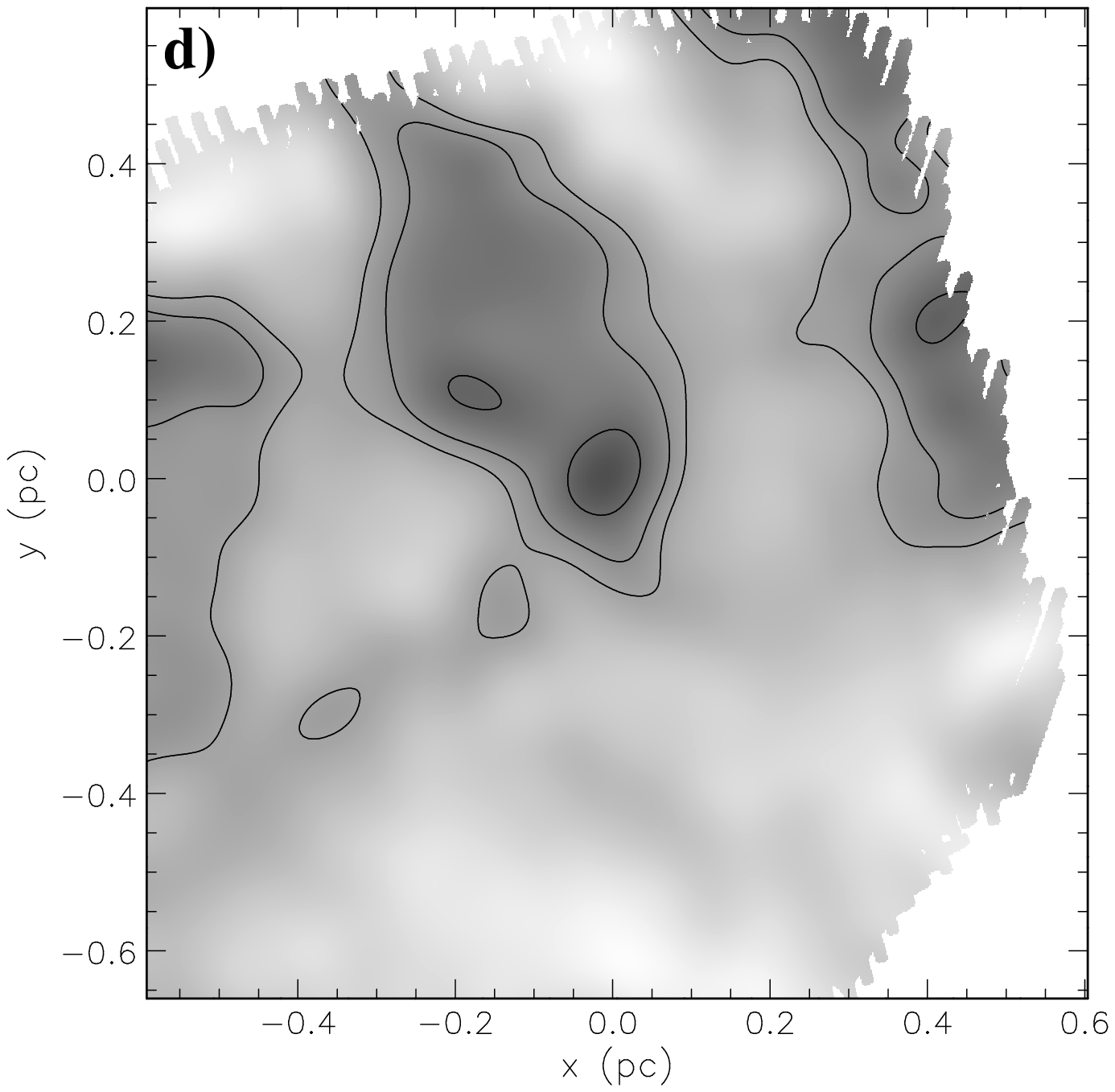}
\end{minipage}
\caption{The best-fit model of L1155C `observed' at (a) 90~$\mu$m and (b) 850~$\mu$m. The heating source is off the bottom of the figure. For comparison, the actual core is shown at (c) 90~$\mu$m and (d) 850~$\mu$m. These data have been rotated to match the observations and the model. The data have also been smoothed to a resolution of 60 arcsec to smooth out the fine details of the core that we do not attempt to model. In each case, the axes show the linear dimensions assuming a distance of 280~pc, with the origin at the centre position of the core.} 
\label{model_rt_maps_1155}
\end{figure*}

The initial shape of the core was taken to be a symmetric sphere, with a plummer-like radial profile \citep{1911MNRAS..71..460P,2001ApJ...547..317W}. This profile is approximately flat at the centre of the core and falls off with a power-law slope to large radii, according to the Plummer relation:
\begin{equation}
n(H_2)(r)=n(H_2)_{flat} \left[ \frac{R_{flat}}{(R_{flat}^2 + r^2)^{1/2} } \right]^\eta,
\label{plummer}
\end{equation}
where $r$ is the distance from the centre of the core, $n(H_2)_{flat}$ is the central density, $R_{flat}$ is the extent of the core in which the density is approximately uniform, and $\eta$ is the power-law slope at large values of $r$. A Plummer-like profile was selected because it provides a high-density inner region, which decreases to large radii, with a small number of parameters. This initial profile was used to model the density profile of pre-stellar cores by \citet[][]{2001ApJ...547..317W}. We use the term Plummer-like because a true Plummer profile has $\eta=5$, whereas we treat this as a free parameter. 

The core is divided into a number of cells by spherical and conical surfaces. There are 70 spherical surfaces which are evenly spaced in radius, and 30 conical surfaces which are evenly spaced in polar angle. Hence the core is divided into 2100 cells. The number of cells used is chosen so that the density and temperature differences between adjacent cells are small.

The value of $R_{flat}$ was determined from the \citet[][]{2005PASJ...57S...1D} extinction map of L1155C. Ideally, we would use the sub-mm data to estimate the extent of the core. However, the chopped nature of the sub-mm data means that they are only sensitive to spatial scales up to approximately two-times the largest chop throw of 1 arcmin. This makes these data unsuitable for measuring the large-scale extent of the core. The extinction data however, are not based on chopped observations, and are also unaffected by the dust temperature. Therefore they trace the large-scale material well.

An azimuthally averaged column density profile was first calculated from the extinction map. Equation \ref{plummer} was then used to generate a volume density profile, which was decomposed to a column density profile under the assumption of spherical symmetry. $R_{flat}$ in equation \ref{plummer} was varied to match the observed and derived column density profiles. The results of this are shown in Figure~\ref{density_prof} for an $R_{flat}$ of 0.35~pc.

The central density ($n(H_2)_{flat}$) was calculated from the peak flux-density of the SCUBA 850~$\mu$m data, which traces the densest region of the core, and is only minimally affected by the temperature gradient. The 6 arcmin resolution of the \citet[][]{2005PASJ...57S...1D} extinction data make these data unsuitable for probing the very centre of the L1155C core, and hence for determining the central density. The uncertainty in the column density derived from the 850~$\mu$m data is approximately $\pm75\%$, taking into account the uncertainties in the distance, dust temperature, and dust opacity. However, when this is converted to a volume density, the unknown physical size of the core along the line of sight introduces a further uncertainty that is difficult to quantify. However, the aspect ratio of the core on the plane of the sky is approximately unity, so it is likely that the 3-dimensional aspect ratio is also of order unity.

The parameters of the initial model, derived as described above, are $R_{flat}=0.35$~pc, and $n(H_2)_{flat} = 2\times 10^4$ cm$^{-3}$. $\eta$ was determined to be equal to 2, based on the fitting of the model to the data.

\begin{figure}
\includegraphics[angle=270,width=87mm]{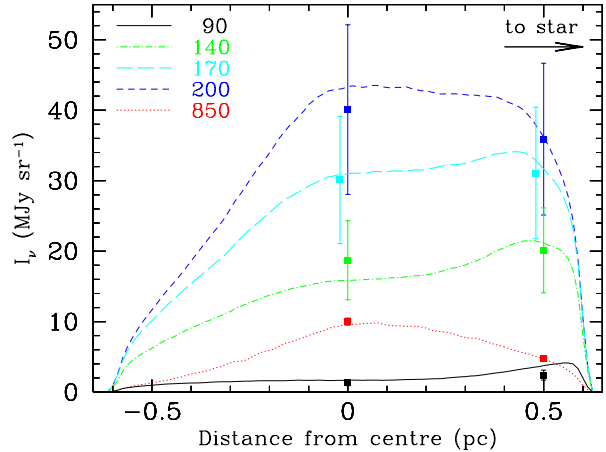} 
\caption{Graph of the emission from the best-fit model of L1155C at each observed wavelength, as a function of distance from the heating source, measured along a straight line connecting the core and the heating source. The flux densities at both the centre and the front positions are plotted.} 
\label{rad_profs_1155}
\end{figure}               

However this model was unable to produce a good fit to the data. To improve the fit, the density profile was skewed towards the front of the core (i.e. towards the heating source). The form of this non-spherical Plummer-like profile is given by:
\begin{equation}
n(H_2)(r,\theta) = n(H_2)_{flat}\,\frac{1 + A \left( \frac{r}{R_{flat}} \right)^2
{\rm sin}^p(\theta/2) }{ \left[ 1 + \left(
\frac{r}{R_{flat}} \right)^2 \right]^\eta} \,
\end{equation}
(Stamatellos et al. 2004). The density profile for this model is shown in Figure~\ref{density_map}.  The azimuthal angle $\theta$ is defined with its origin pointing away from the heating source (the $+y$ direction in Figure \ref{density_map}). The parameter $A$ determines the "south-to-north" optical depth ratio $e\,$, i.e. the maximum optical depth from the centre to the surface of the core (which occurs at $\theta = 180\degr$), divided by the minimum optical depth from the centre to the surface of the core (which occurs at $\theta = 0\degr$). The parameter $p$ determines how rapidly the optical depth rises with increasing $\theta$, from the north  at $\theta = 0\degr$ to the south at $\theta = 180\degr$. For the best-fit simulation we use $R_{flat}=0.35$~pc, $n(H_2)_{flat} = 2\times 10^4$ cm$^{-3}$, $p=4$, $A=56.07$, $e=1.5$ and $\eta=2$. Due to these additional free parameters, we accept that this model is not a unique solution. However, the amount of skewing required is a relatively minor perturbation on an accepted basic pre-stellar density profile \citep{2001ApJ...547..317W}.

\begin{figure}
\includegraphics[angle=270,width=87mm]{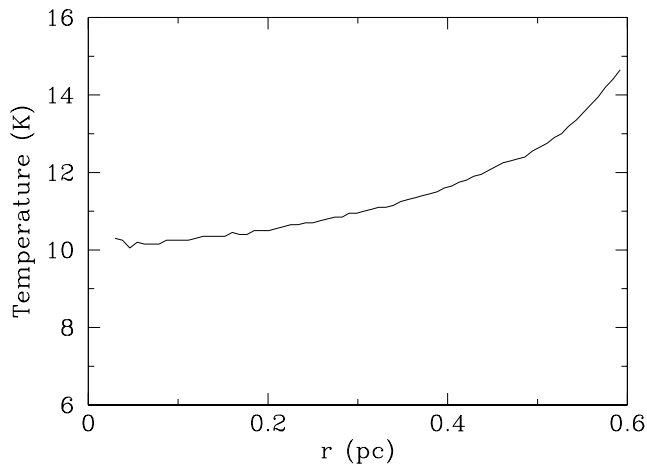} 
\caption{The temperature of the best-fit model of L1155C, plotted as a function of the distance from the centre of the core in the direction of the heating source. Note the increase in temperature towards the front of the core.}
\label{temp_prof_1155}
\end{figure}               

We also note, that this model is designed to reproduce the large-scale appearance of the L1155C core at each of the observed wavelengths. The model does not attempt to model any of the substructure of L1155C that is apparent in, for example, the 850~$\mu$m data. 

The results of the radiative transfer modelling are shown in Figures~\ref{model_rt_maps_1155},~\ref{rad_profs_1155} and~\ref{temp_prof_1155}. Figure~\ref{model_rt_maps_1155} shows the model core `observed' at (a) 90 and (b) 850~$\mu$m using the {\sc PHAETHON} code. For comparison, the data for the same two wavelengths are shown in Figure~\ref{model_rt_maps_1155} (c) and (d). These data have been rotated to highlight the similarities with the model, and a qualitative agreement between the model and the data at both wavelengths is clear.

Figure~\ref{rad_profs_1155} quantifies the similarity between the model and the observations by taking a 1-D cut through the model core at each wavelength. Each line shows the emission from the model core as a function of distance from the heating source, measured along a line connecting the core and the heating source. 

The flux densities at the centre and the front positions (listed in Table~\ref{aperture_fluxes}) are also plotted on this figure for comparison. There is very good consistency between the observations and the model at all wavelengths.

\begin{figure*}
\begin{minipage}{87mm}
\hspace{5mm}
\includegraphics[angle=0,width=68.8mm]{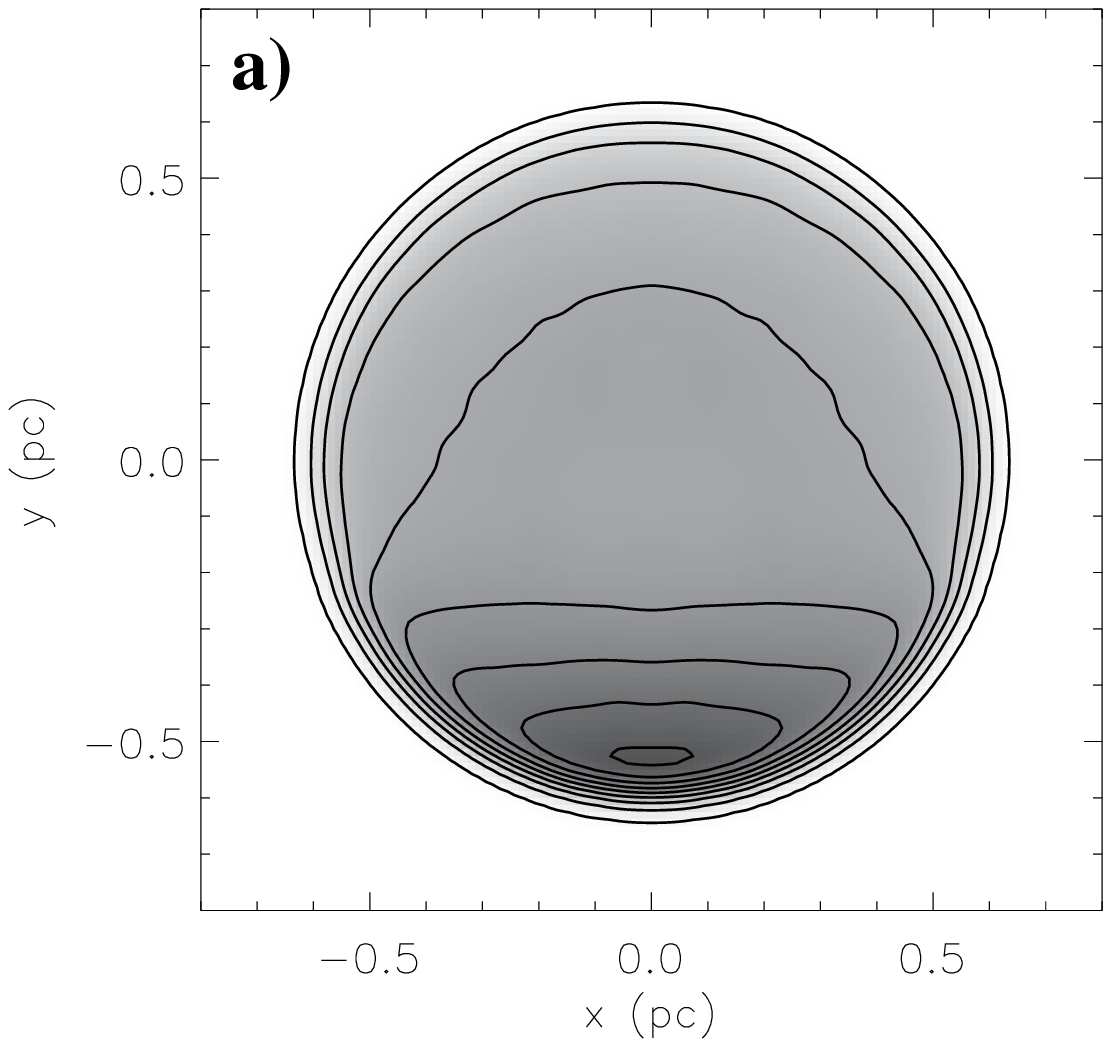} 
\end{minipage}
\begin{minipage}{87mm}
\hspace{-5mm}
\includegraphics[angle=0,width=68.8mm]{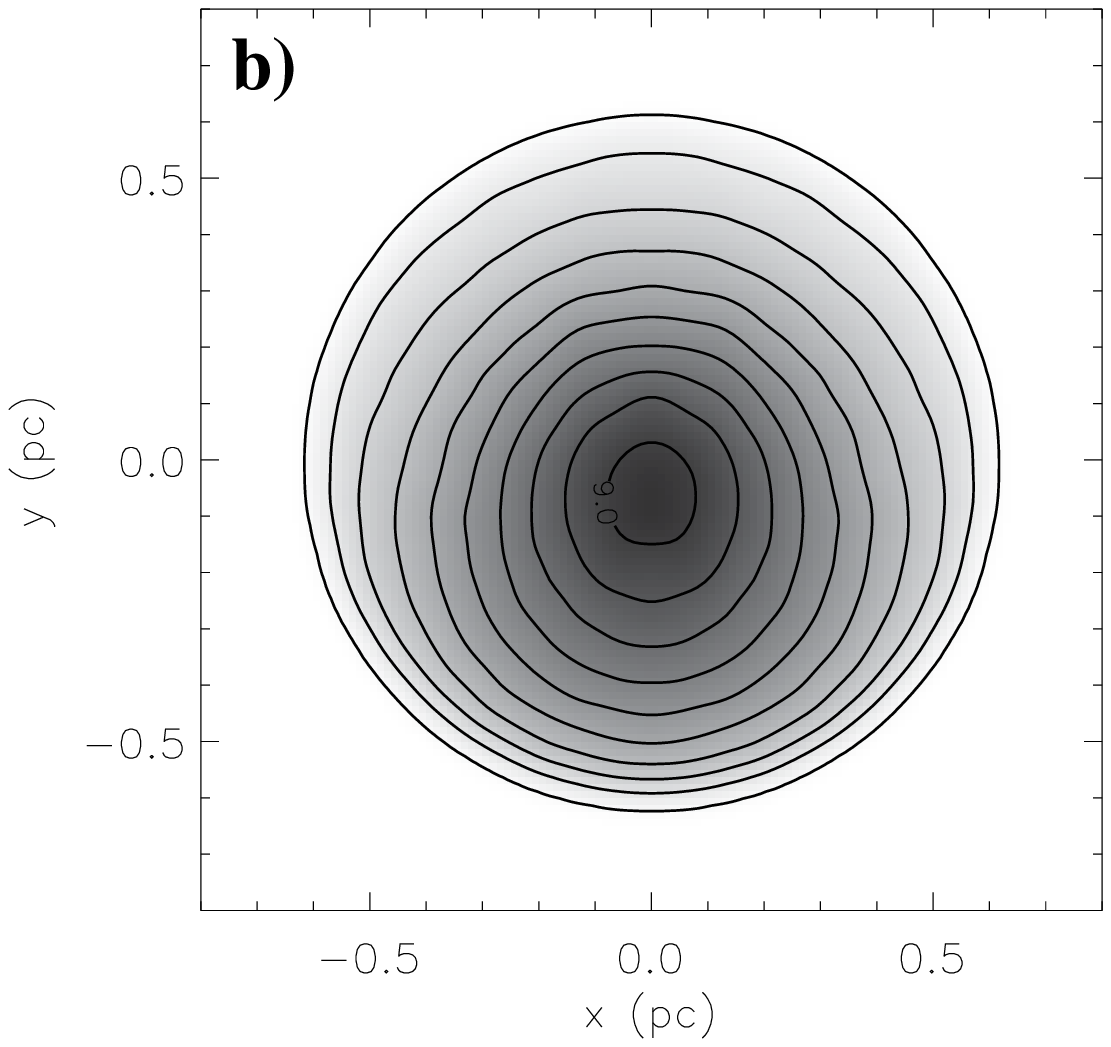}
\end{minipage}
\begin{minipage}{87mm}
\hspace{8.5mm}
\includegraphics[angle=0,width=65mm]{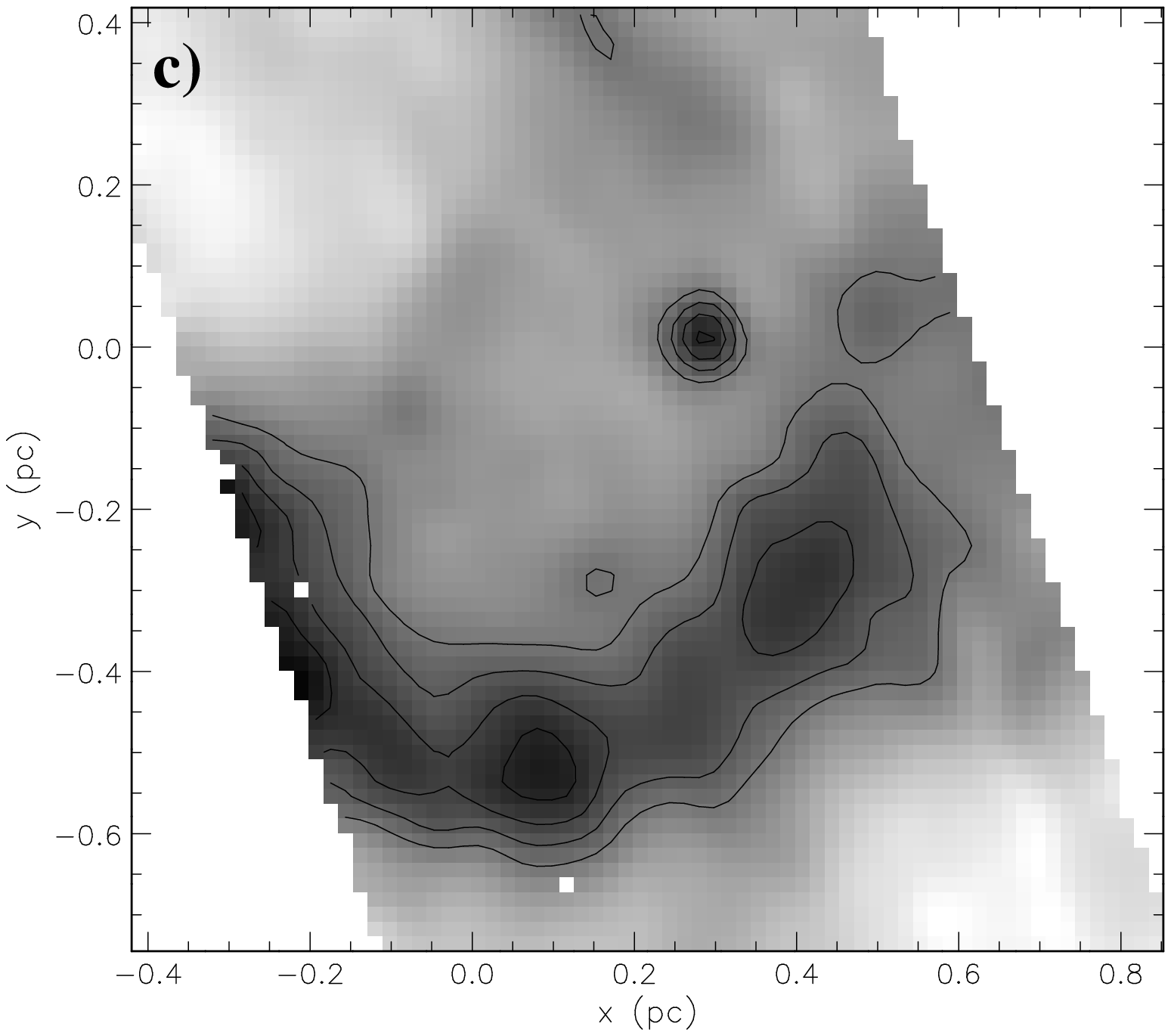} 
\end{minipage}
\begin{minipage}{87mm}
\hspace{-1.5mm}
\includegraphics[angle=0,width=65mm]{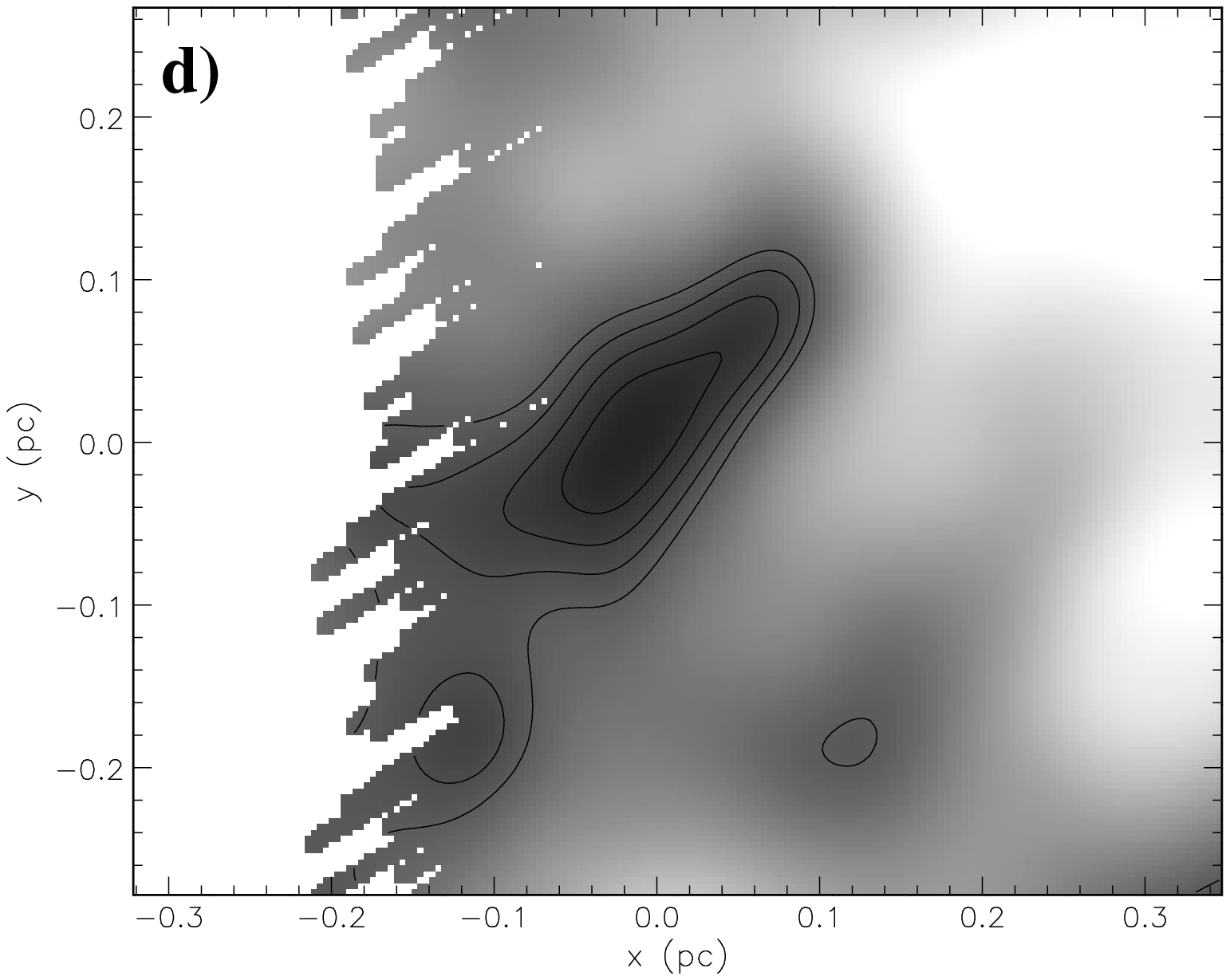}
\end{minipage}
\caption{The model of L1148 `observed' at (a) 90~$\mu$m and (b) 850~$\mu$m. The heating source is off the bottom of the figure. For comparison, the actual core is shown at (c) 90~$\mu$m and (d) 850~$\mu$m. As for Figure \ref{model_rt_maps_1155}, the data have been rotated to match the observations and the model, and smoothed to a resolution of 60 arcsec. In each case, the axes show the linear dimensions assuming a distance of 250~pc, with the origin at the centre position of the core.} 
\label{model_rt_maps_1148}
\end{figure*}

The temperature profile of the best-fit model is shown in Figure~\ref{temp_prof_1155}. The temperature rises from $\sim 10$~K in the centre of the core to $\sim$13~K at the edge of the core nearest to the heating source. These temperatures are lower than we measured using the SED fits, especially at the central position of the core. However, this is expected in an externally heated core because the observations sample warmer material in front of the core as well as the colder material at the centre \citep{2007MNRAS.379.1390S}. 

We note that a change in the properties of the emitting dust grains might also affect the appearance of the core at different wavelengths, but the shift in grain properties would have to be dramatic, and a modest temperature variation seems far more likely.

To conclude our study of the L1155C core, our hypothesis of a temperature gradient causing the observed change in core morphology at different wavelengths is confirmed by the SED fitting of the data at two positions in the core. In addition, a radiative transfer model of L1155C with a geometry based on observed parameters shows that this temperature gradient can realistically be caused by a nearby bright star. This model successfully reproduces both the observed core morphology and measured flux density at each wavelength, and also the observed temperature gradient.

\subsection{Radiative Transfer Modelling - L1148}\label{rt-1148}

The L1148 core contains the very low luminosity Class 0 (VeLLO) L1148-IRS, and therefore represents an intermediate phase between a pre-stellar core, and a protostellar core. The protostar has formed, but its luminosity is not yet strongly influencing the core.

The L1148 core exhibits a similar morphology to the L1155C core in the Akari data, with the far-infrared emission wrapping around the peak of the extinction map (see Figures \ref{fig_akari90}, \ref{fig_akari140} and \ref{fig_akari160}). We hypothesise that this is caused by the same process as in L1155C and that the observed morphology in the far-infrared is being affected by a temperature gradient. 

The data that we have available to constrain the model of this core are the 90, 140 and 160~$\mu$m Akari data over the whole core, and 850~$\mu$m data of the centre of the core only. The 850~$\mu$m data for the centre of L1148 are shown in Figure \ref{model_rt_maps_1148}d, and show that the core is centrally condensed, and peaks somewhat behind the crescent shaped 90~$\mu$m peak (see Figure \ref{model_rt_maps_1148}c), as is the case in L1155C.

These data allow us to fit a modified blackbody curve to the SED at the centre position of L1148, as we did for L1155C in Section \ref{temp}. The best-fit blackbody has a temperature of 11.5~K, similar to the centre of L1155C. However, the lack of a long wavelength data-point for the front position means that we cannot constrain the blackbody at this position. Hence we cannot establish the presence or absence of a temperature gradient in L1148 other than by analogy of the similar morphologies of L1148 and L1155C at far-infrared and sub-mm wavelengths.

Figure \ref{colour_composite} shows that the star BD+67 1263 is not in a suitable location to heat the L1148 core where we see it most brightly at 90~$\mu$m. A more likely heating source is the star HD197053 (see Table \ref{stars} for details). This star is located to the west of the L1148 core and has a high heating flux of between 300 and 600${\rm \times 10^{-9} Wm^2}$ at the front of the core.

We modelled the emission from the L1148 core with the {\sc PHAETHON} radiative transfer code, using the same geometry as the best-fit model for L1155C. Again, the extent of the core and the peak density were determined using the \citet{2005PASJ...57S...1D} extinction data, and the 850~$\mu$m data respectively.

\begin{figure}
\includegraphics[angle=270,width=87mm]{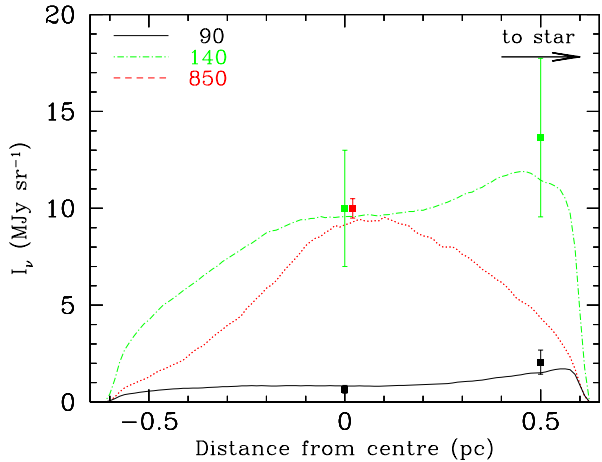} 
\caption{Graph of the emission from the model of L1148 at each observed wavelength, as a function of distance from the heating source, measured along a straight line connecting the core and the heating source. The flux densities at both the centre and the front positions are plotted.} 
\label{rad_profs_1148}
\end{figure}

The results of this model are shown in Figures \ref{model_rt_maps_1148} and \ref{rad_profs_1148}. Figure \ref{model_rt_maps_1148} shows that there is a qualitative agreement between the model and the data. As for L1155C, the model reproduces the centrally condensed emission at 850~$\mu$m, and the region of strong emission at the front of the core at 90~$\mu$m. The temperature of the model core increases from $\sim$~10~K at the centre, to $\sim$~13~K at the front of the core. Again, the central temperature in the model is expected to be lower than the measured value. 

The agreement between the model and the data is shown quantitatively in Figure \ref{rad_profs_1148}. This shows the emission from the core as a function of distance from the centre in the direction of the heating source. The model is compared to the measured flux densities at the centre and front positions of L1148, and shows a good agreement.

In conclusion, though we have fewer data for the L1148 core, the data are consistent with the core being externally heated. This heating is dramatically influencing the appearance of the core at far-infrared wavelengths. This is especially interesting, because L1148 harbours the very low luminosity Class 0 L1148-IRS \citep{2005AN....326..878K}. However, the data are consistent with the core being heated externally rather than there being a temperature gradient caused by this heating source. This is possibly due to the very low luminosity of this source.

\section{Conclusions}\label{conclusions}

We have presented far-infrared Akari data of the L1155 region at 90, 140 and 160~$\mu$m. We have compared these data to data of similar wavelengths as observed by the ISO satellite, and also to sub-mm data taken with SCUBA on the JCMT. The Akari data are shown to be consistent with the ISO data, although the Akari data cover a larger area. 

The data span the wavelength range of 90 -- 850~$\mu$m, and have a monotonic shift in peak position of L1155C as a function of wavelength. We interpret this as a temperature gradient across the core. This hypothesis is backed up with SED fitting at the centre of the core and also at the front of the core, where the short-wavelength emission peaks. The best-fit temperature for the centre of L1155C is 11.9~K, compared to a temperature of 13.6~K at the front edge of the core.

We carried out radiative transfer modelling of the L1155C core using the PHAETHON code. The code subjects a model core to the interstellar radiation field, plus a component from the nearby A6V star BD+67 1263. The core is treated as a plummer-like sphere, which has been skewed slightly towards the radiation source. With this model, we reproduce the appearance of the core, as observed at 90 -- 850~$\mu$m. We also reproduce both the temperature gradient, and the absolute temperature values.

The L1148 core exhibits a similar morphology at far-infrared and sub-mm wavelengths. Though we have fewer data available for this core, we are able to fit these data with a similar plummer-like sphere, which is externally heated from the nearby B-type star HD197053.

The data presented here demonstrate that far-infrared data can very useful for interpreting longer wavelength data. In particular it can be useful for determining the temperature of cores, information which cannot be derived from sub-mm data alone. However, we note that far-infrared data should be used with caution for studying pre-stellar cores when complementary sub-mm data are not available. This study has shown how the density structure is not immediately apparent from the far-infrared data alone, and the sub-mm data are required to obtain a complete picture of the core. This needs to be borne in mind when studying data from other far-infrared satellites such as Spitzer and Herschel.

\section*{Acknowledgments}
The work is based on observations with AKARI, a JAXA project with the participation of ESA. 

The James Clerk Maxwell Telescope is operated by The Joint Astronomy Centre on behalf of the Science and Technology Facilities Council of the United Kingdom, the Netherlands Organisation for Scientific Research, and the National Research Council of Canada. The observations presented here were taken under the programs M02AN18 and M03BN10.

The Infrared Space Observatory is an ESA project with instruments funded by ESA member states particularly the PI countries, France, Germany, Netherlands and the United Kingdom and with the participation of ISAS (Japan) and NASA (USA). 

The Digitized Sky Survey was produced at the Space Telescope Science Institute under U.S. Government grant NAG W-2166. The images of these surveys are based on photographic data obtained using the Oschin Schmidt Telescope on Palomar Mountain and the UK Schmidt Telescope at Siding Spring. The plates were processed into the present compressed digital form at the Royal Edinburgh Observatory (ROE) photolabs with the permission of these institutions. 

The authors acknowledge the use of NASA's SkyView facility (http://skyview.gsfc.nasa.gov) located at NASA Goddard Space Flight Center. 

This research made use of Montage, funded by the National Aeronautics and Space Administration's Earth Science Technology Office, Computation Technologies Project, under Cooperative Agreement Number NCC5-626 between NASA and the California Institute of Technology. Montage is maintained by the NASA/IPAC Infrared Science Archive.

The authors wish to thank J. Kirk for the use of his SED fitting scripts, and Nicholas Chapman for the use of his Spitzer-IRAC extinction map of L1155C.

DN and DS acknowledge STFC for PDRA support under the Cardiff University Astronomy Rolling Grant.

\end{document}